\documentclass{aa}  
\usepackage{graphicx}
\usepackage{caption}
\usepackage{nccmath}
\usepackage[varg]{txfonts}
\usepackage{hyperref}
\hypersetup{
colorlinks=true,
citecolor=blue
}

\newcommand{\rcs}{RCS2\,032727$-$132623}
\newcommand{\pks}{PSZ1\,G311.65$-$18.48}
\newcommand{\mgii}{\ion{Mg}{ii}}
\newcommand{\sgas}{SGAS\,J1226+2152}
\newcommand{\siii}{\ion{Si}{ii}}
\newcommand{\cii}{\ion{C}{ii}}

\begin{document} 

   %\date{Received September 15, 1996; accepted March 16, 1997}
   \title{Directly constraining the spatial coherence of the $z\sim1$ circumgalactic medium}
   %\subtitle{}

   \author{A. Afruni\inst{1}\and S. Lopez\inst{1}\and
           P. Anshul\inst{1}
           \and
           N. Tejos\inst{2}\and
            P. Noterdaeme\inst{3,4}\and
            T. A. M. Berg\inst{1,5,6}\and 
            C. Ledoux\inst{5}\and
            M. Solimano\inst{7}\and\\ 
            J. Gonzalez-Lopez\inst{7}\and
            M. Gronke\inst{8}\and
            F. Barrientos\inst{9}\and
            E. J. Johnston\inst{7}
          }

   \institute{Departamento de Astronom\'ia, Universidad de Chile, Camino el Observatorio 1515, Las Condes, Santiago, Chile\\
              \email{aafruni@das.uchile.cl}
    \and
    Instituto de F\'isica, Pontificia Universidad Cat\'olica de Valpara\'iso, Casilla 4059, Valpara\'iso, Chile
    \and
    Franco-Chilean Laboratory for Astronomy, IRL 3386, CNRS and Universidad de Chile, Santiago, Chile
    \and
    Institute d'Astrophysique de Paris, CNRS-SU, UMR 7095, 98bis bd Arago, 75014 Paris, France
    \and
    European Southern Observatory, Alonso de Cordova 3107, Casilla 19001, Santiago, Chile
    \and
    Dipartimento di Fisica G. Occhialini, Universit\`a degli Studi di Milano Bicocca, Piazza della Scienza 3, 20126 Milano, Italy
    \and
    Instituto de Estudios Astrof\'isicos, Facultad de Ingenier\'ia y Ciencias, Universidad Diego Portales, Av. Ej\'ercito Libertador 441, Santiago, Chile
    \and
    Max Planck Institut f\"ur Astrophysik, Karl-Schwarzschild-Stra{\ss}e 1, D-85748 Garching bei M\"unchen, Germany
    \and
    Instituto de F\'isica, Pontificia Universidad Cat\'olica de Chile, Av. Vicu\~na Mackenna 4860, 7820436 Macul, Santiago, Chile.
             }

  \abstract
  { One of the biggest puzzles regarding the circumgalactic medium (CGM) is the structure of its cool ($T\sim10^4$ K) gas phase. While the kinematics of quasar absorption systems suggests the CGM is composed of a population of different clouds, constraining the clouds' extent and spatial distribution has proven challenging, both from the theoretical and observational points of view. In this work we study the spatial structure of the $z\sim 1$ CGM with unprecedented detail via resolved spectroscopy of giant gravitational arcs. We put together a sample of \mgii$\lambda\lambda 2796,2803$ detections obtained with VLT/MUSE in 91 spatially independent and contiguous sight-lines toward 3 arcs, each probing an isolated star-forming galaxy believed to be detected in absorption. We constrain the coherence scale of this gas ($C_{\rm{length}}$), which represents the spatial scale over which the \mgii\ equivalent width (EW) remains constant, by comparing EW variations measured across all sight-lines with empirical models. We find $1.4 <C_{\rm{length}}/\rm{kpc} <7.8$ (95\% confidence). This measurement, of unprecedented accuracy, represents the scale over which the cool gas tends to cluster in separate structures. We argue that, if $C_{\rm{length}}$ is a universal property of the CGM, it needs to be reproduced by current and future theoretical models in order to  understand the exact role of this medium in galaxy evolution.}
    \titlerunning{Constraining spatial coherence of CGM}
   \maketitle
%
%________________________________________________________________

\section{Introduction}\label{intro}
It is now clear that the evolution of galaxies must be linked to their interaction with their surrounding gaseous halos, the circumgalactic medium, or CGM. Of particular importance is the CGM cool phase \citep[$T\sim10^4$ K, see][]{tumlinson17}, which traces the flows of material in and out of galaxies, known as the baryon cycle, and represents a huge reservoir of fuel for star-formation.
However, granted that this medium holds the key to understand how galaxies grow and evolve, many of its properties are to date still unclear.

Due to its low emissivity, observing the cool CGM directly in emission is challenging and currently most of these observations are focused on the high redshift Universe ($z\gtrsim2$), where there seems to be an ubiquity of extended Ly$\alpha$ nebulae, illuminated by the central quasars or galaxies \citep[e.g.,][]{cantalupo14,farina19}. At $z\la1$, there are instead only a handful of cases where the cool CGM has been observed in emission, either around starbursts and galaxy groups \citep[e.g.,][]{chen19groupNebula,burchett21,leclercq22} or thanks to stacking techniques \citep[e.g.,][]{zhang18}. The vast majority of the cool CGM detections come then from absorption studies, extensively performed for decades both at high and low redshift, using as tracers the absorption lines of hydrogen and low-ionization ion species like \mgii, \cii\ or \siii\ in the spectra of background quasars \citep[e.g.,][]{bergeron86,churchill00,zhu13,keeney17,zahedy19,Qu23}. 

A major issue in understanding the elusive properties of this  material is the presence of typically only one single line of sight per galaxy, so that the properties of the cool CGM are usually studied in a statistical way, combining together detections from surveys of tens or hundreds of galaxy-QSO pairs \citep[e.g.,][]{lanzetta95,chen10,werk12,wilde21}. This has proven very useful in unveiling some of the properties of this medium, like its absorption strength or its covering fraction as a function of the distance from the central galaxy \citep[e.g.,][]{nielsen13,huang21}. However, due to the single-pencil-beam limitation, there are still very strong uncertainties on the general dynamics and structure of this gas. High spectral resolution observations have shown that this medium is likely composed of multiple clumps of gas, or clouds, which produce multiple kinematic components along the same line of sight and which are bound to the galaxy halo \citep[e.g.,][]{borthakur15}, but whether these clouds are primarily inflowing toward \citep[e.g.,][]{bouche13} or outflowing \citep[e.g.,][]{schroetter19} from the central galaxies is still a matter of debate. 

From a theoretical perspective, the cool CGM properties across the galaxy halo can be studied using cosmological and zoom-in simulations. However, given the low resolution in the CGM (kpc-scale), such works are not yet converged in terms of the cool gas dynamics and structure \citep[e.g.,][]{vandevoort19} and the picture is to date uncertain.  
On the other hand, high-resolution (pc-scale) hydrodynamical simulations, focused on the motion of a single cool cloud interacting with the surrounding hotter CGM phase ($T\sim10^6$ K), have shown that the dynamics and the survival \citep[e.g.,][]{marinacci10, gronke20} of such clouds highly depend on their initial size and mass, as found also by analytical models \citep[e.g.,][]{nipoti07,afruni19,fielding22}. Constraining these sizes is therefore crucial to advance our understanding of the cool CGM dynamics and its connection with the central galaxy. But estimating these values directly from observations has proven extremely challenging.

Several studies have resorted to photoionization models like CLOUDY \citep[e.g.,][]{ferland98} to estimate the thickness of the cool CGM clouds~\citep[e.g.,][]{werk14}, but these estimates are subject to large  uncertainties (up to several orders of magnitude) due to the inherent assumptions of the models~\citep{haehnelt96}. A more direct way to estimate the spatial extension of these clouds is to use multiply lensed quasars \citep[e.g.,][]{smette95,lopez99,ellison04,lopez07,chen14,zahedy16,rubin18lensedQSO}, which allow one to probe the absorbers through more than just one line of sight. However, the results on absorber sizes are inconclusive due to the small number of bright lenses and the small number (2 or 3) of lines of sight per system, preventing a broader view of the gas structure. 

The present work is largely motivated by the promising approach of using extended background sources~\citep{steidel10,bordoloi14,rubin18}. Using a sample of 27 galaxy-galaxy pairs taken from the PRIMUS survey \citep{coil11}, \cite{rubin18} estimated the scale over which the absorption strength of the cool CGM, traced by \mgii\ absorption, is expected to vary. Through their analysis, they were able to put a lower limit on this quantity at about $2$ kpc. This means that they did not observe variations in the absorption strength on scales smaller than this threshold and that even if the single clouds have smaller sizes, their distribution and kinematics result in a constant absorption strength on scales of at least a few kpc.  
This coherence scale gives therefore an idea of what is the clustering scale of the cool medium. In this work we aim at estimating this quantity with  greater accuracy, through the use of gravitational arcs. 

In recent years, the ARCTOMO collaboration\footnote{\tt
  https://sites.google.com/view/arctomo}  has developed a new technique to spatially resolve the cool CGM in absorption across individual galaxy halos. 
  The technique, hereafter "arc-tomography", 
  uses resolved spectroscopy of giant gravitational arcs as background sources, observed with integral-field spectrographs like VLT/MUSE~\citep{bacon10}. The observations enable studies of the cool CGM distribution and kinematics (as traced by \mgii\ absorption) of single galaxies \citep[e.g., ][]{lopez18,lopez20,tejos21,ferfig22} in an unprecedented fashion by directly mapping its properties across the arc. In the present work, we use this unique dataset (hereafter "ARCTOMO data"), combined with the predictions from empirical models, to constrain the cool CGM coherence scale, defined in the same way as in \cite{rubin18}. 

In Sect.~\ref{data}, we summarize our data, describing the sample selection, data reduction, and absorption line search and we define the main observational constraints that we will utilize in this work; in Sect.~\ref{method}, we outline the creation of our fiducial CGM model and the technique used to compare the model outputs with the data; in Sect.~\ref{results}, we report our findings regarding the cool CGM coherence scale, in Sect.~\ref{discussion} we discuss the interpretation of the coherence scale, the limitations of our analysis, the comparison with previous studies and the implications of our findings for theoretical models, while in Sect.~\ref{conclusions} we summarize the main conclusions of this study.

Throughout the paper, we assume a flat $\Lambda$CDM cosmology with $H_{0}=70$ km s$^{-1}$ Mpc$^{-1}$, $\Omega_{\Lambda} = 0.7$ and $\Omega_{M} = 0.3$.

\section{Observational data}\label{data}
\begin{table*}
  \centering
\caption{ARCTOMO data}
\resizebox{\textwidth}{!}{%
\begin{tabular}{lccccccccccc}
\hline
\hline
Field&\multicolumn{4}{c}{G1 emission properties$^a$}&&\multicolumn{5}{c}{absorption
  properties}& \\
\cline{2-5} \cline{7-11}\\
&$z$ & $L_B/L^*_B$ & $\log(M_*/M_\sun)$ & SFR& & \# sp. & \# det.& EW & $R$ & scale&refs.\\
& & & & [$M_\sun~{\rm yr}^{-1}$] &  & & &[\AA] &[kpc] &[kpc] & \\
&(1)& (2)& (3)& (4) && (5) & (6) & (7)&(8) & (9)& (10)\\
\cline{2-7} \cline{8-12}\\
\pks & 0.73379 & 0.14$\pm$0.03 &9.7$\pm$0.3 &1.1$\pm$0.3 & & 148 & 29 & 0.2 -- 3.7 & 0 -- 43 & 2.8
&L20\\ 
\rcs & 0.98238 & 0.11$\pm$0.01 &9.8$\pm$0.2 & 0.4$\pm$0.1&  & 103 & 26 & 0.3 -- 1.9 & 17 -- 52 & 2.9 & L18\\ 

\sgas & 0.77138 & 0.27$\pm$0.02 &10.1$\pm$0.1 &0.6$\pm$0.2 &  & 90 & 36 & 0.6 -- 4.2 & 1 -- 21 & 2.6 & T21\\ 
\hline
\label{table_data}
\end{tabular}
}
\vspace{-0.5cm}
\flushleft
Notes: 
(1) redshift;
(2) B-band luminosity; 
(3) stellar mass;  
(4) star-formation rate; 
(5) number of $0.6\arcsec\times0.6\arcsec$ binned spaxels with $\rm{S/N}>2$;
(6) number of \mgii\ detections (significance $>$ 2 in both doublet lines);
(7) range in \mgii$\lambda 2796$ rest-frame equivalent width;
(8) range in impact parameters$^b$ in the absorber plane (only detections); 
(9) median linear scale of the binned spaxels (only detections) in the absorber plane;
(10) references: L20 \citet{lopez20}$^a$; L18 \cite{lopez18}; T21 \cite{tejos21}. 
\\
Footnotes: 
$^a$ In each field "G1" is the identified absorbing galaxy. All properties are corrected for magnification. \rcs\ G1 is resolved into 3 galaxies; integrated properties are listed.  
$^b$ Projected distance in the absorber (reconstructed) plane between G1 and spaxels. 
\end{table*}
%%%%%%%%%%%%%%%%%%%%%%%%%%%%%%%%%%%%%%%%%%%%%%%%%%%%%%%%

Table~\ref{table_data} lists summary properties of the 3 systems
studied in this work, that is, the sample of spatially independent
\mgii\ detections and their identified absorbing galaxies (hereafter called `G1').  These systems were found in giant gravitational-arc
fields observed by us with VLT/MUSE. The selection for this work was based on
absorbing galaxies (1) having a large number of spatially independent
\mgii\ detections in their CGM; (2) sharing similar
redshifts, stellar masses, luminosities and star formation rates; and (3) being considered ``isolated''. The
``isolation'' character holds strictly only for G1 toward \pks, where
no galaxies at the same redshift ($\Delta v\la 1\,000\ \rm{km}\ \rm{s}^{-1}$) are found in the MUSE
field~\citep{lopez20}, while for G1 toward \sgas\ one galaxy at the
same redshift is found over $225$ kpc away (de-lensed projected
distance)~\citep{tejos21} and for G1 toward \rcs\ two galaxies are
found at the same redshift  over 200 kpc away~\citep{lopez18}. In addition \rcs\ G1 itself is resolved into 3
galaxies in the {\it HST} images~\citep{lopez18} and hence integrated properties are provided in Table~\ref{table_data}.

Although these systems have been
presented individually in the indicated references, the present work
includes improved data reduction and a re-analysis of the absorption
line profiles, about which we provide details in the following.

\subsection{MUSE observations and data reduction}

VLT/MUSE integral-field observations of these 3 ARCTOMO fields, \pks,
\rcs, and \sgas, were carried out in service mode within ESO programs
297.A-5012(A) (PI Aghanim), 098.A-0459(A) (PI Lopez), and
0101.A-0364(A) (PI Lopez), respectively. All observations were carried
out in wide-field mode, which provides a field-of-view of
$\approx1\arcmin\times1\arcmin$, spaxel size of 0.2\arcsec, and
spectra of resolving power $R=2\,000$--$4\,000$ in the range
$4\,650$--$9\,300$ \AA, at a dispersion of $1.25$ \AA. In addition, \sgas\ was
observed using the Adaptive-Optics extended wavelength mode.

Observations proceeded during dark time and under good weather
conditions.  To minimize instrumental
and sky artifacts, short (640--700 s) individual exposures were taken having small
spatial ditherings and cumulative 90$^\circ$ rotations. The number of exposures
and the total integration times varied according to arc brightness,
resulting in relatively homogeneous continuum signal-to-noise (S/N) levels near the
relevant \mgii\ absorption.  
For more details on the individual observing
conditions we refer the reader to the references shown in
Table~\ref{table_data}.

Data of all 3 fields were reduced in an homogeneous fashion using the
MUSE pipeline~\citep{weilbacher12} using the EsoRex (ESO Recipe Execution
Tool) tool. The wavelength solution
was calibrated to vacuum directly from the pipeline. The effective PSF FWHM in the combined cubes ranges between 0.7\arcsec\ and 0.8\arcsec. Residual sky
contamination was removed using the Zurich Atmosphere Purge
code~\citep[ZAP v2.1,][]{soto16}, using the default parameters, on the final combined cubes.

\subsection{Automated absorption line analysis}\label{absorptionFits}

To obtain spatially resolved rest-frame equivalent width maps, the
absorption line analysis proceeds automatically, in two basic stages:
(i) extraction of arc spectra and (ii) absorption line fitting. For each data
cube the specific tasks are as follows.

We first rebin the cube spatially.  
This task serves two purposes:
account for the cross-talk between neighbor MUSE native spaxels due to
the seeing, and increase the S/N. In a sense this
is equivalent to extracting spectra of many resolved sources in a
coherent fashion. In this work we use ``$3 \times 3$ binning'',
meaning that each binned cube spaxel is $0.6\arcsec$ on a side and
contains a weighted average spectrum out of 9 native spaxels. This choice maximizes spatial sampling while still keeping seeing-induced cross-correlation low between neighbor spaxels \citep{tejos21}.  
We then proceed to select binned spaxels in a
masked region that encompasses the arc based on S/N. The S/N ratio is
calculated in a featureless spectral region close to the expected
\mgii\ absorption using the actual flux RMS. We use $\rm{S/N} =2$ as a
threshold. This limit is arbitrary; it serves the purpose of pre-selecting arc spectra while ruling out spaxels that are  contaminated by the sky specially at the arc borders. The final \mgii\ selection used in this work is done at the absorption-line fitting level.  

To obtain equivalent widths, the spectra are normalized and 
two Gaussians that account for the \mgii\ $\lambda\lambda 2796,2803$ doublet are fitted. Describing the profiles with Gaussians assumes the absorption signal is dominated by the instrumental profile. Thus,  the 4 free
parameters are 2 amplitudes\footnote{The final combined equivalent width distribution shows a cutoff at $\approx 0.3$ \AA\ (see Fig. \ref{fig:Completeness}), which indicates that our spectra do not select individual \mgii\ clouds but rather their unresolved line-of-sight kinematics, which is likely dominated by saturated lines. This justifies the use of double Gaussians with untied amplitudes.}, 1 common line width and 1 common
redshift. Fits are considered successful if the equivalent width measurement is larger than 2 times its 1-$\sigma$ error in both
doublet lines. 
The scheme yields 91 detections among the three fields. The fitted parameters are reported in Appendix~\ref{fits}.
The present work is based on the (rest-frame) equivalent width of the 2796 doublet line, hereafter called EW.   
EWs are assigned to the binned spaxels to create the maps shown in Fig.~\ref{fig:EW_maps_orig} (both in the image and the absorber planes). Only successful fits are considered.   Also, since we are using an updated selection criterion, these maps are different from those in~\citet{lopez18,lopez20} and~\citep{ferfig22}. 

\subsection{Transverse distances in the absorber plane}

\begin{figure}[!h]
    \centering
   \includegraphics[clip, trim={0cm 0cm 0cm 0cm}, width=0.9\columnwidth]{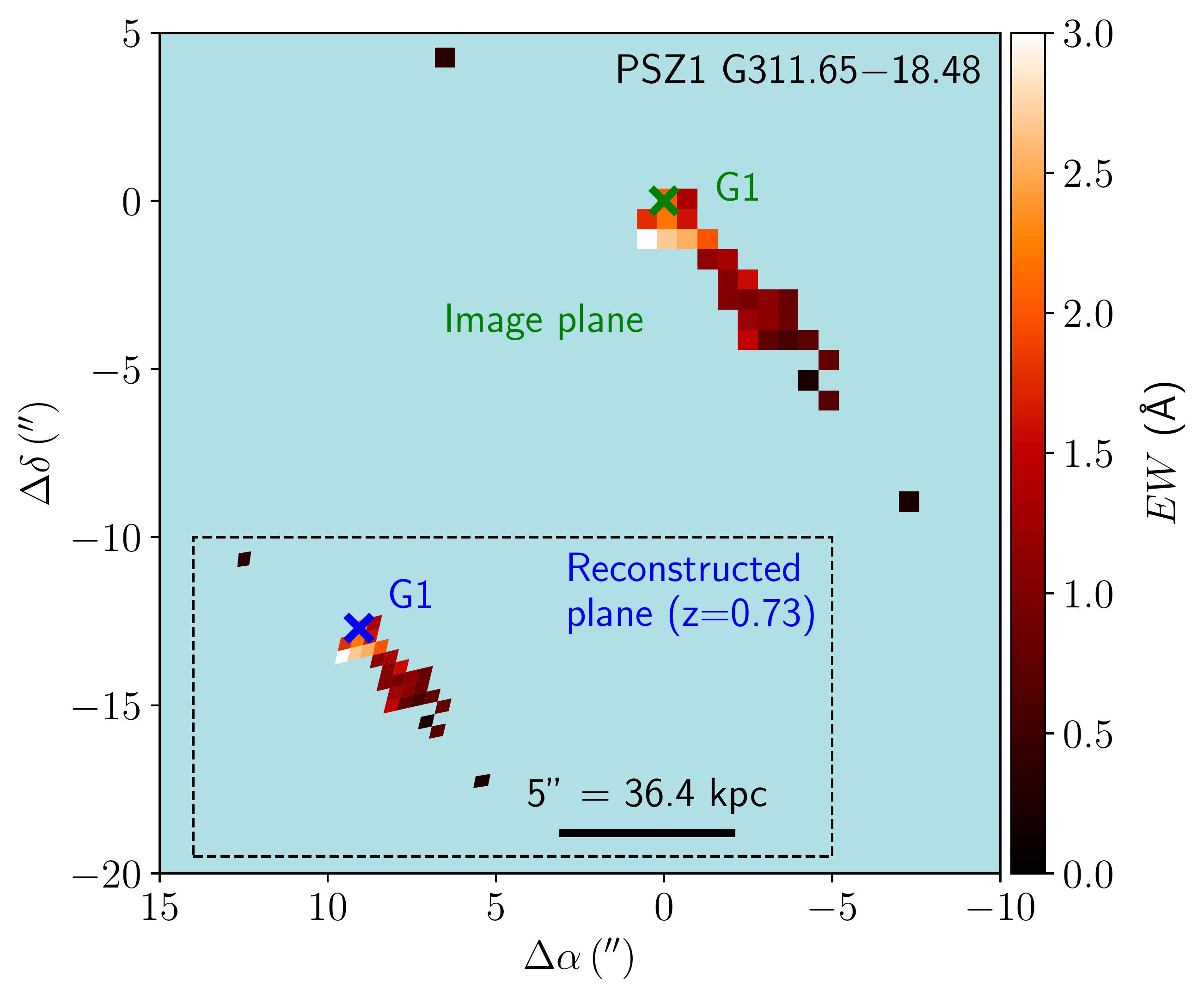}
   \includegraphics[clip, trim={0cm 0cm 0cm 0cm}, width=0.9\columnwidth]{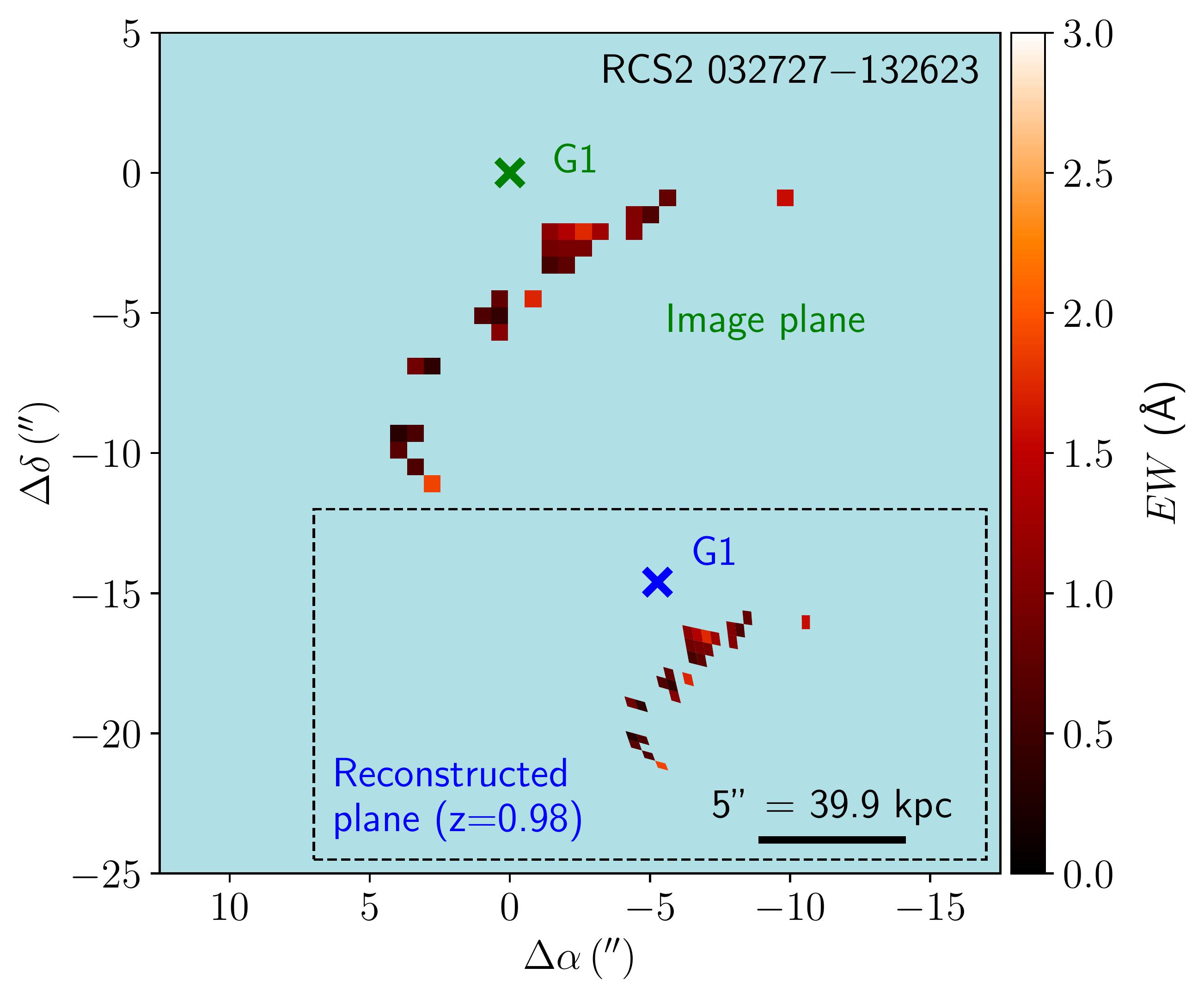}
   \includegraphics[clip, trim={0cm 0cm 0cm 0cm}, width=0.9\columnwidth]{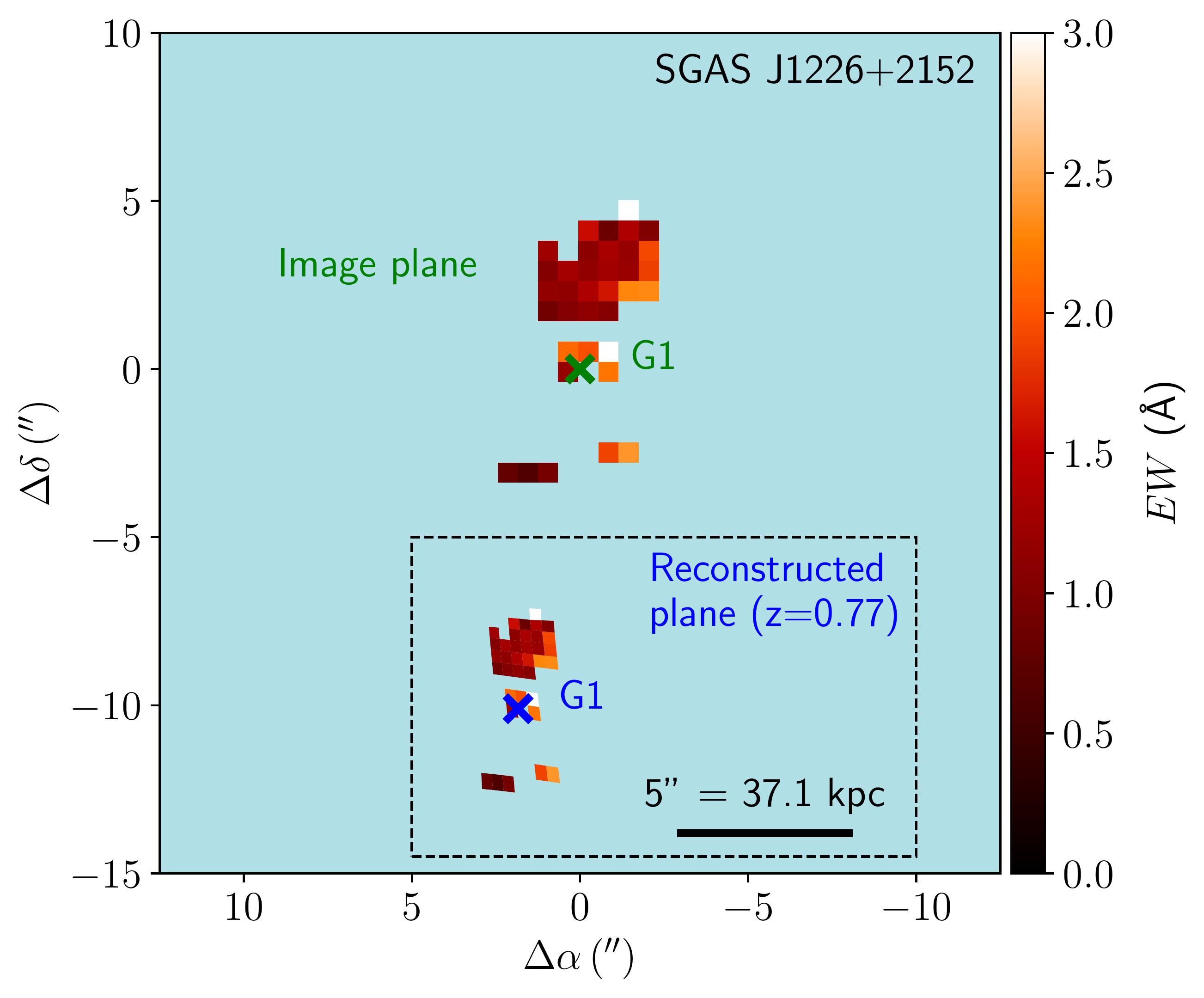}
   \caption{Maps of equivalent widths of the 2976 \mgii\ doublet line, for all the detections found in the three fields analyzed in this work: \pks\ (top), \rcs\ (center) and \sgas\ (bottom). Both the image (observed) plane and the reconstructed absorber plane (dashed-line rectangle) are shown, with the crosses showing the positions of the three identified absorbing galaxies, named G1. In \sgas\ the absorption on top of G1 is shown for completeness but not used in the analysis.}
              \label{fig:EW_maps_orig}%
\end{figure}

At the absorber redshifts all RA-DEC coordinates are affected by lensing. 
We transform those to  "de-lensed" coordinates using the lens models
described in detail in the references listed in
Table~\ref{table_data}. Such a 
transformation produces a map of de-lensed spaxels in the absorber plane (also called "re-constructed" plane, see Fig.~\ref{fig:EW_maps_orig}) at the redshift of the absorber. The de-lensed 
coordinates are used to measure projected physical separations  between spaxel centers and also between spaxels and the central galaxy. We call this latter quantity "impact parameter", or $R$.

Lens models are precise to around 5\% \citep{lopez20} and this uncertainty propagates to $R$. We have checked that such uncertainty is however negligible for the purposes of the present study and does not affect our findings. But the models are also subject to systematic uncertainties depending on how well they are constrained observationally, so the solutions might not
be unique. For instance, our model for the \sgas\ field considers two
perturbers while the model presented in \cite{mortensen21} just one,
resulting in two different reconstructed absorber planes for this
field. However, the two different re-constructed spaxel maps can be reasonably well matched with each other by a translation of coordinates only.  
Hence, $R$ and transverse distances between spaxels are unlikely to be strongly affected by systematic uncertainties in the lens model \citep{tejos21}.
\subsection{Main observational constraints}\label{mainObs}
   \begin{figure*}
   \includegraphics[clip, trim={0cm 0cm 0cm 0cm}, width=0.49\linewidth]{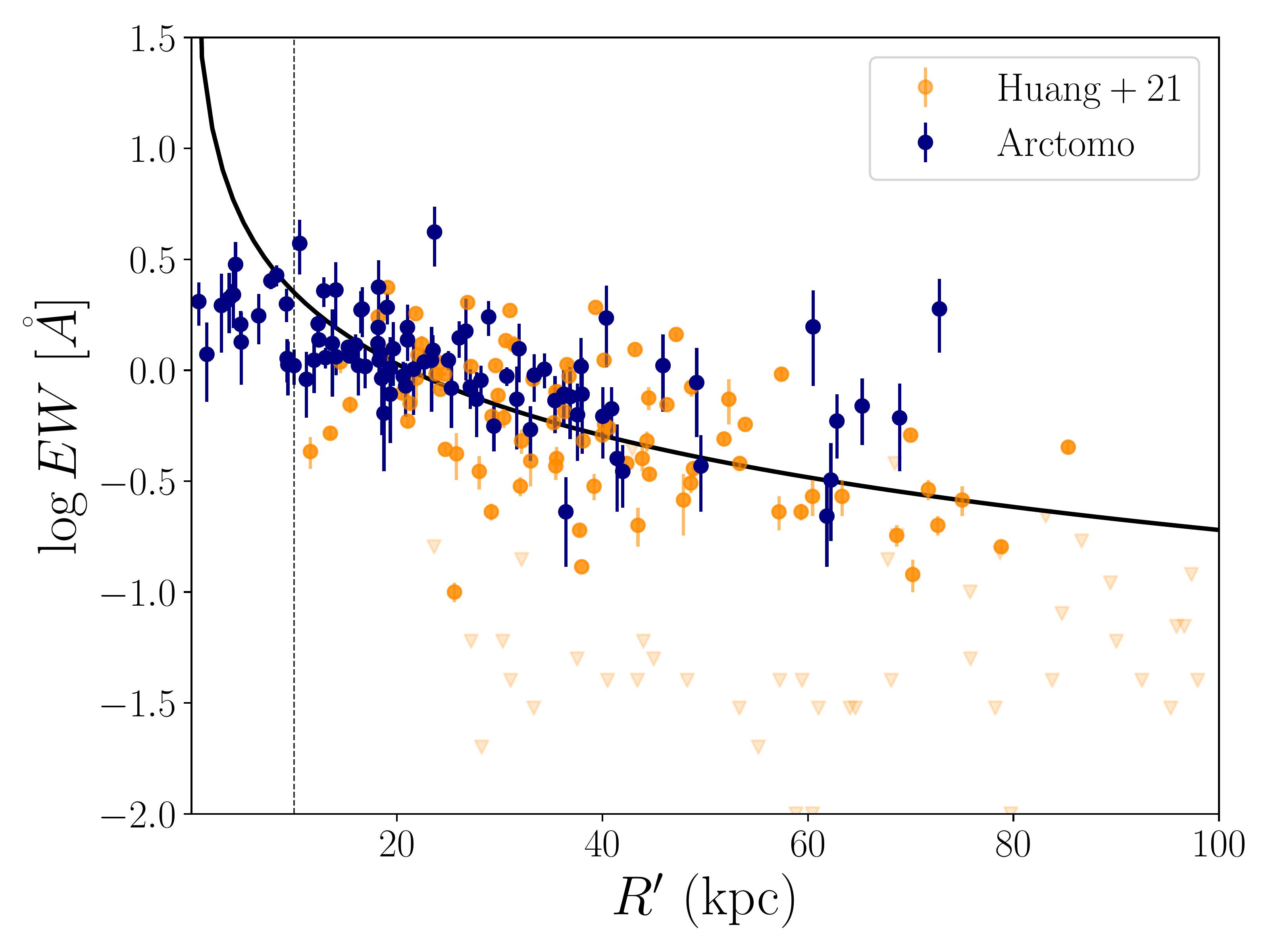}
   \includegraphics[clip,  
   trim={0cm 0cm 0cm 0cm},
   width=0.49\linewidth]{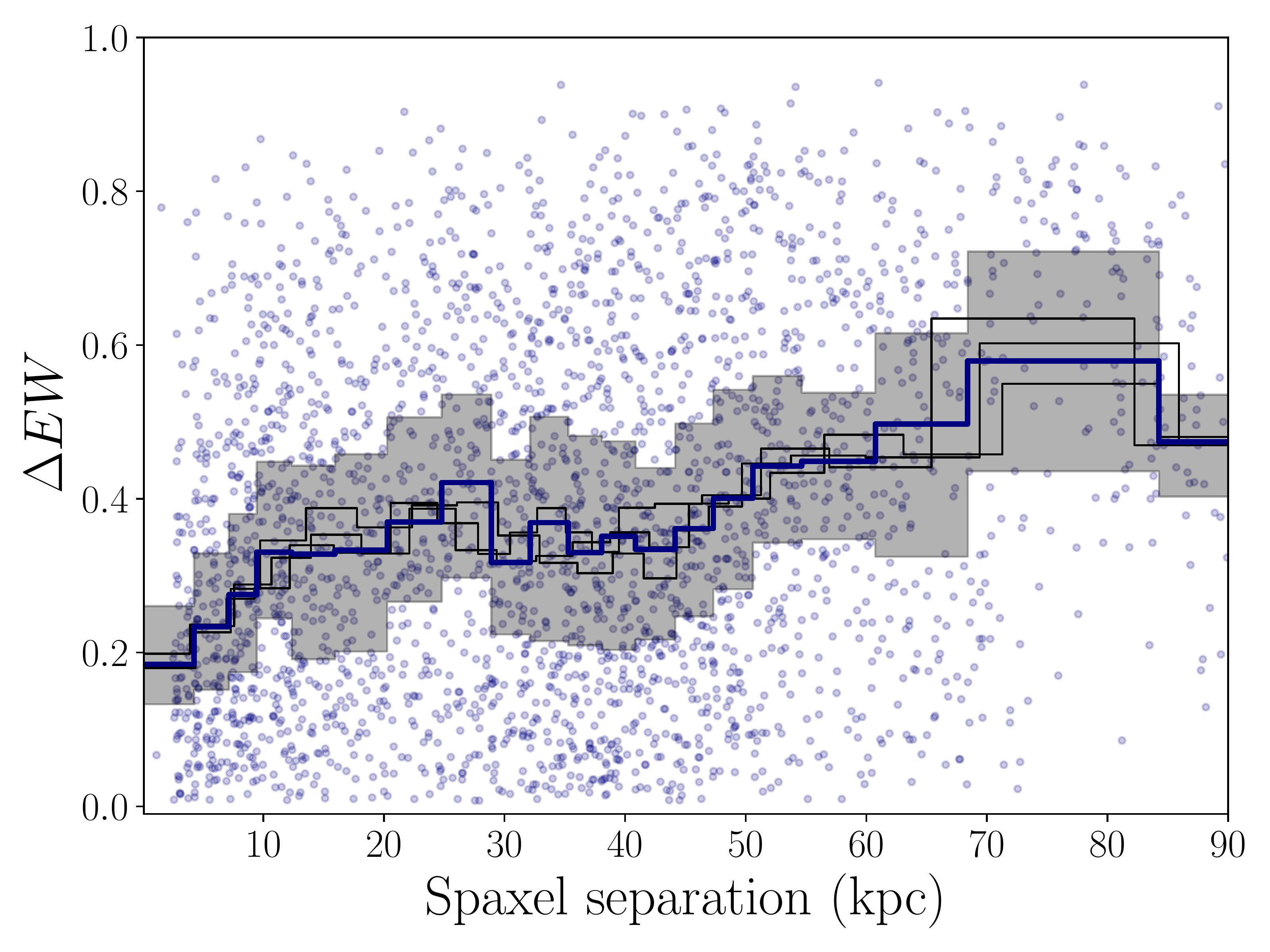}
   \caption{Main observational constraints extracted from the ARCTOMO data. Left: \mgii$\lambda2796$ rest-frame equivalent width as a function of $R'$ (Eq.~\ref{eq:Rcorr}). In blue, arctomographic data, in orange, data from \cite{huang21}, with 2-$\sigma$ upper limits denoted by downwards triangles. The black curve shows the best-fit of the \cite{huang21} sample (Eq.~\ref{eq:wR}) and the vertical dashed line marks the region where our data tend to diverge from this curve (these data are excluded from our analysis). Right: distribution of the fractional difference of EW in the arctomo data, as a function of the spaxel separation. The blue points show the values of all the spaxel pairs, while the blue curve and the gray band show the median values, with the 1-$\sigma$ uncertainties, of a binning of the same distribution (each bin contains 150 values). The black curves show the results for distributions obtained by choosing different orientation angles for the three MUSE fields (see Sect.~\ref{mainObs} for more details).}
              \label{fig:EWvsRbins}%
    \end{figure*}

In this work we combine 
EWs and spaxel positions with respect to the central galaxy of the three fields described above. The underlying assumption is that the cool CGM has self-similar properties across our sample of galaxies. 
This is a common practice in absorption studies of the CGM, but  while it is normally assumed for samples of tens or even hundreds of galaxies, here we focus instead on the halos of only three galaxies, with similar properties (see Table~\ref{table_data}). To take into account the slight differences in the stellar masses of the three galaxies, we define a re-scaled version of $R$, as: 
\begin{ceqn}
\begin{equation}\label{eq:Rcorr}
R' = R / (M_{\ast}/M_{\ast,0})^a\ ,
\end{equation}
\end{ceqn}
where $M_{\ast,0}=10^{10.3}\ M_{\odot}$ and $a=0.26$. This definition was introduced in \cite{huang21}, who conducted a survey with MagE of \mgii\ absorbers around almost 400 galaxies at $z<0.4$ using more than 150 quasars as background sources. In particular, the values of $M_{\ast,0}$ and $a$ reported above are derived from fitting the \mgii\ EW as a function of the impact parameter and the stellar mass (see below) for their sample of isolated star-forming galaxies, similar to the objects analyzed in the present study. We then investigate how the \mgii\ equivalent width varies in our data as a function of $R'$ and as a function of the spaxel separation, as described below. We emphasize that, for simplicity, we use only confirmed \mgii\ detections. We consider that most of non-detections are actually due to spectra that (despite our selection based on S/N, see Sect.~\ref{absorptionFits}) are at the arc borders and therefore contaminated by the sky. We plan to address this information in future works, as it cannot be trivially implemented in our models (see Sect.~\ref{method}). 

In the left panel of Fig.~\ref{fig:EWvsRbins}, we show in blue the \mgii\ EWs of the three ARCTOMO fields with their respective 1-$\sigma$ uncertainties, all as a function of $R'$: we can see that the EW decreases with increasing the distance from the central galaxy. This distribution represents the first observational constraint that will be used in this study. As a comparison, we also report, in orange, the data coming from the survey of \cite{huang21} (considering only the sample of isolated star-forming galaxies), with the corresponding best-fit curve in black, given by:
\begin{ceqn}
\begin{equation}\label{eq:wR}
\log EW = a_0 + a_1\log R'\ ,
\end{equation}
\end{ceqn}
where $a_0=1.42, a_1 = -1.07$. The ARCTOMO data lie on top of those obtained from the quasar survey and are well-described by the quasar-based best-fit relation. The main difference can be seen in the region where $R'\lesssim10$ kpc, where  however there are no data available from the \cite{huang21} sample and the best-fit relation represents only an extrapolation. All the data with $R'\lesssim10$ kpc are therefore excluded from our analysis (see Appendix~\ref{fits}).

The second main observable that we use to constrain the coherence length of the cool CGM is given by the distribution of the EW fractional differences, called $\Delta EW$, as a function of spaxel separation. We define the fractional EW difference between two spaxels as \citep[e.g.,][]{ellison04,rubin18lensedQSO,okoshi21}:

\begin{ceqn}
\begin{equation}\label{eq:fracEW}
\Delta EW = \frac{|EW_1 - EW_2|}{\max{(EW_1, EW_2)}}\ .
\end{equation}
\end{ceqn}

\noindent We calculate this quantity for all the pairs of spaxels with  detections 
and combining together the three MUSE fields.  
The error on $\Delta EW$ is propagated from the uncertainties of the individual EWs (assuming that the errors of $EW_1$ and $EW_2$ are uncorrelated). We then create a distribution of $\Delta EW$ as a function of spaxel separation. 
To calculate the distances between spaxels located in different MUSE fields, we assume random orientation angles between the fields. A Monte Carlo simulation shows that these relative orientations do not affect the distribution of $\Delta EW$ vs. separations. One of these realizations is shown on the right-hand panel of Fig.~\ref{fig:EWvsRbins} (uncertainties are not shown for clarity).  
We bin separations by equal number of pairs (150) and compute the median $\Delta EW$ (blue solid line) and the 32nd and 68th percentiles (gray band) in each bin. 
We can appreciate how $\Delta EW$ is on average already larger than 0 for contiguous spaxels and tends to increase with  spaxel separation. Note that the peculiar shape of this distribution for separations between 30 and 60 kpc is likely not physical, but due to the specific spaxel configuration we are using (a similar shape is indeed visible also in the synthetic observations of our models, Fig.~\ref{fig:fracEWComparison}). In this work, we will anyway focus on the small spaxel separations.

Finally, the three additional black curves in the right-hand-side of Fig.~\ref{fig:EWvsRbins} show the median values for three other distributions with different choices of the field orientations: these lines are all within the gray band, showing how, as mentioned above, the orientation does not affect this distribution.  This suggests that the EW distribution is isotropic, such that it does not show any increment or decrement of EW in a particular direction with respect to the host galaxy. 

\section{Models}\label{method}
\begin{figure}[!h]
    \includegraphics[width=\linewidth]{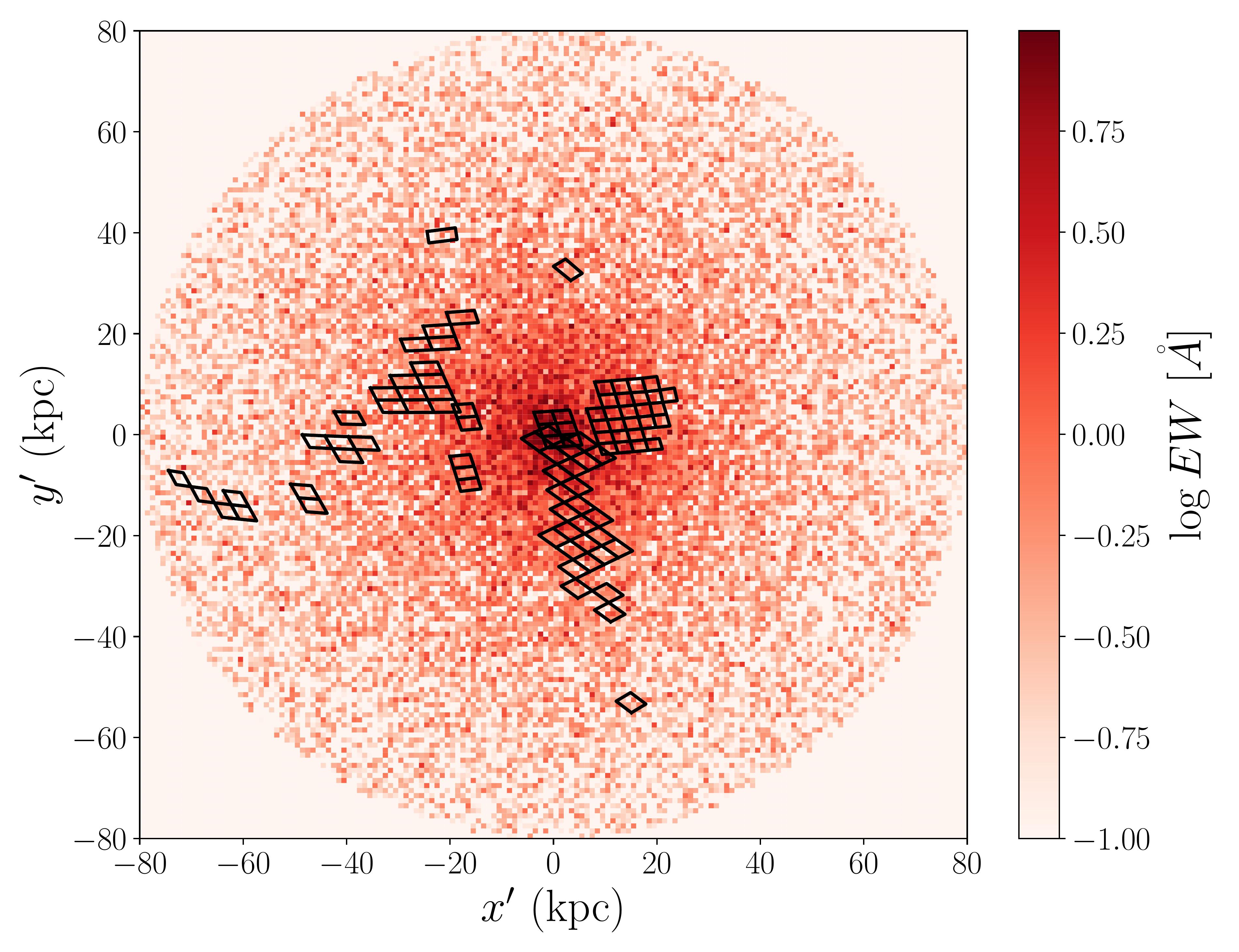}
   \includegraphics[ width=\linewidth]{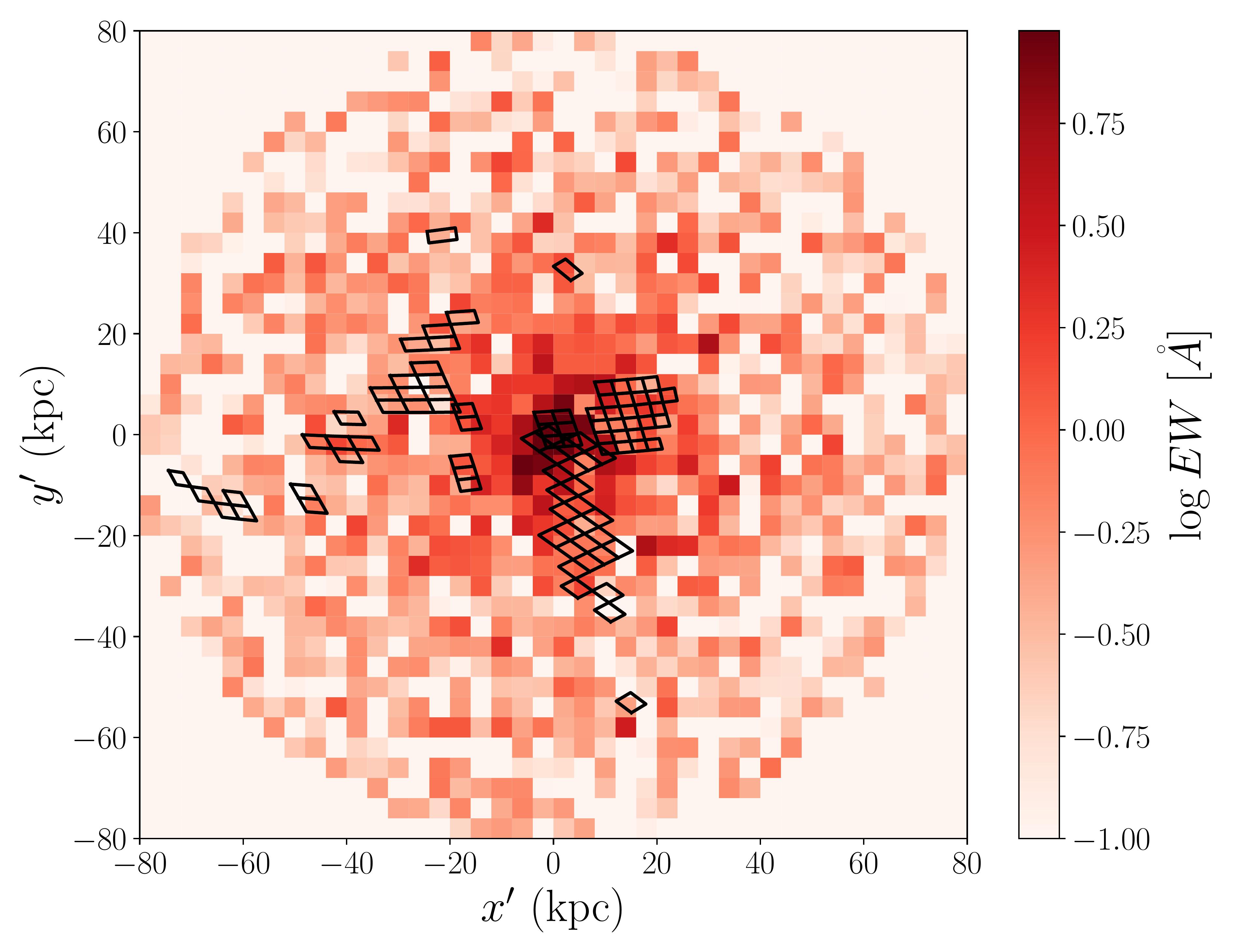}
   \includegraphics[ width=\linewidth]{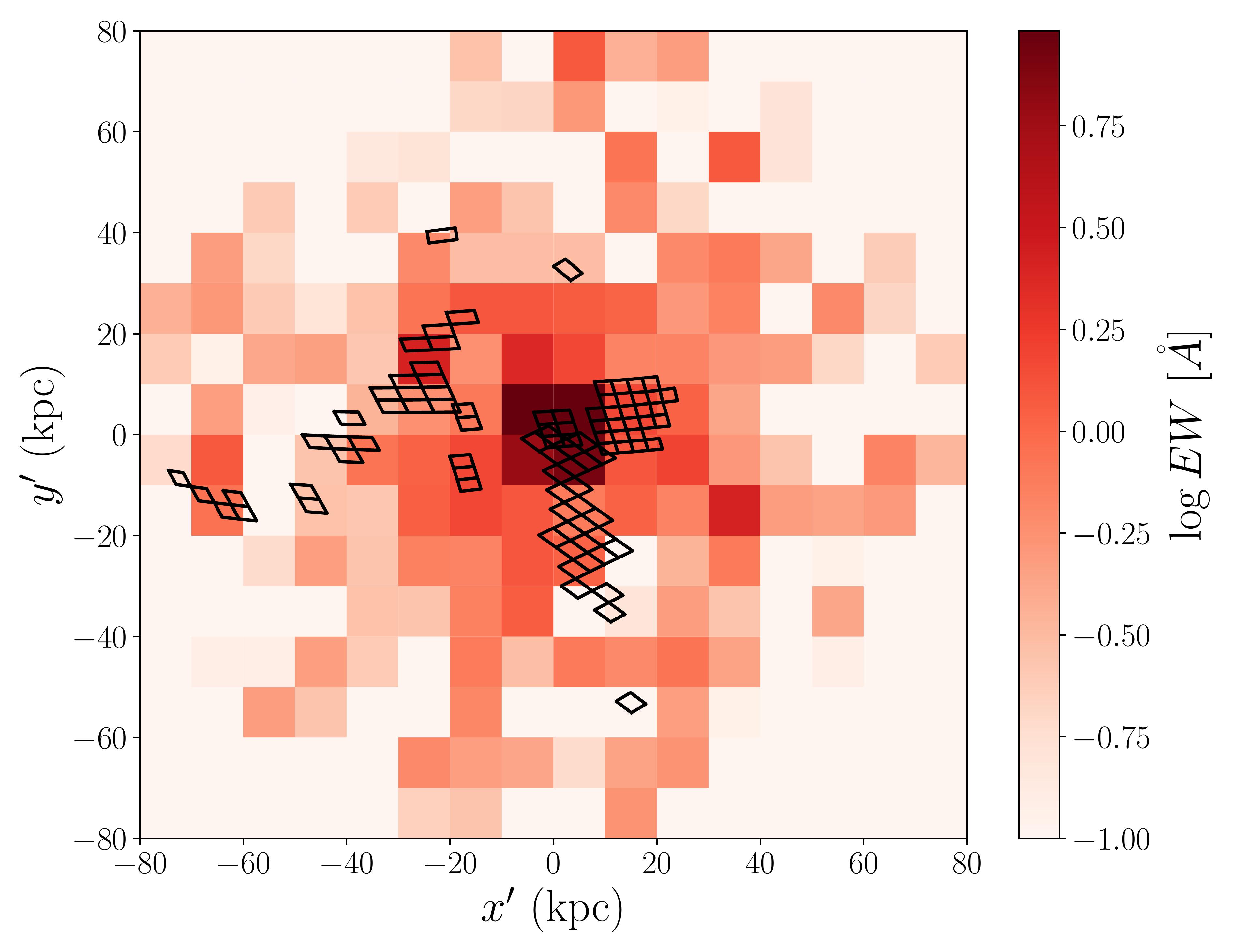}
   \caption{EW maps obtained using the method described in Sect.~\ref{mapCreation}, for a coherence length of 1 kpc (top), 4 kpc (center) and 10 kpc (bottom). The spaxel distribution used to create the synthetic observations (Sect.~\ref{syntheticObs}) is superimposed in black.}
              \label{fig:EWmaps}%
    \end{figure}
\subsection{Basic approach}
To interpret our observables, we build empirical models of the cool CGM distribution.  
We create 2-dimensional maps of \mgii\ EW, where a crucial parameter is given by the gas coherence length, $C_{\rm{length}}$. As in \cite{rubin18}, we define the coherence length (or scale) as the size over which the equivalent width of \mgii\ remains constant. In practice, our maps are divided in cells, whose size is equal to $C_{\rm{length}}$: the EW cannot vary within a cell, while different cells can have different EW values.

The two main assumptions when creating our models are i) the `intrinsic' distribution of EWs across the halo follows the results of the quasar survey from \cite{huang21}, justified by the fact that our data are consistent with their findings; ii) the EW distribution is isotropic across the plane, justified by the distribution of fractional EW differences in our data (see Sect.~\ref{mainObs} and Fig.~\ref{fig:EWvsRbins}). 
We then perform synthetic observations, tracing MUSE-like spaxels that resemble the observations outlined in Sect.~\ref{data} and assuming different values for the coherence scale. The outputs of such artificial observations can be directly compared with the observational constraints outlined in Sect.~\ref{mainObs}, allowing us to estimate which choice of $C_{\rm{length}}$ better reproduces the ARCTOMO data.
\subsection{Creation of EW maps}\label{mapCreation}
We populate the re-scaled (Eq.~\ref{eq:Rcorr}) projected plane $(x', y')$ surrounding the central galaxy (located at $x'=0,\ y'=0$), using the EW-vs-$R'$ relation described by Eq.~\eqref{eq:wR} and a declining covering fraction profile based on the constraints found by \cite{huang21}\footnote{We use a power-law profile obtained by fitting the values reported in Table 4 of \cite{huang21}.}. 
In detail, we first divide the plane into a grid with a cell-size equal to the assumed coherence length and then into concentric rings with a width equal to the coherence length itself; we then populate each ring  
with a number of `systems' $n_{\rm{ring}}$ equal to the total number of cells in the ring multiplied for the predicted covering fraction. This is done by assigning to the selected $n_{\rm{ring}}$ cells a value of EW taken from a Gaussian distribution centered on the mean value of EW predicted by Eq.~\eqref{eq:wR} for those distances and with a width equal to the intrinsic scatter $\sigma$ of the data from \cite{huang21}. This scatter is equal to 0.278 for $R'<100$ kpc, which is the region analyzed in this paper. Finally, the `empty' cells have EW $=0\ \AA$\footnote{Choosing an arbitrary low value instead of 0 does not impact the results of this paper.}.

   \begin{figure*}[!h]
   \includegraphics[clip, trim={0cm 0cm 0cm 0cm}, width=0.95\linewidth]{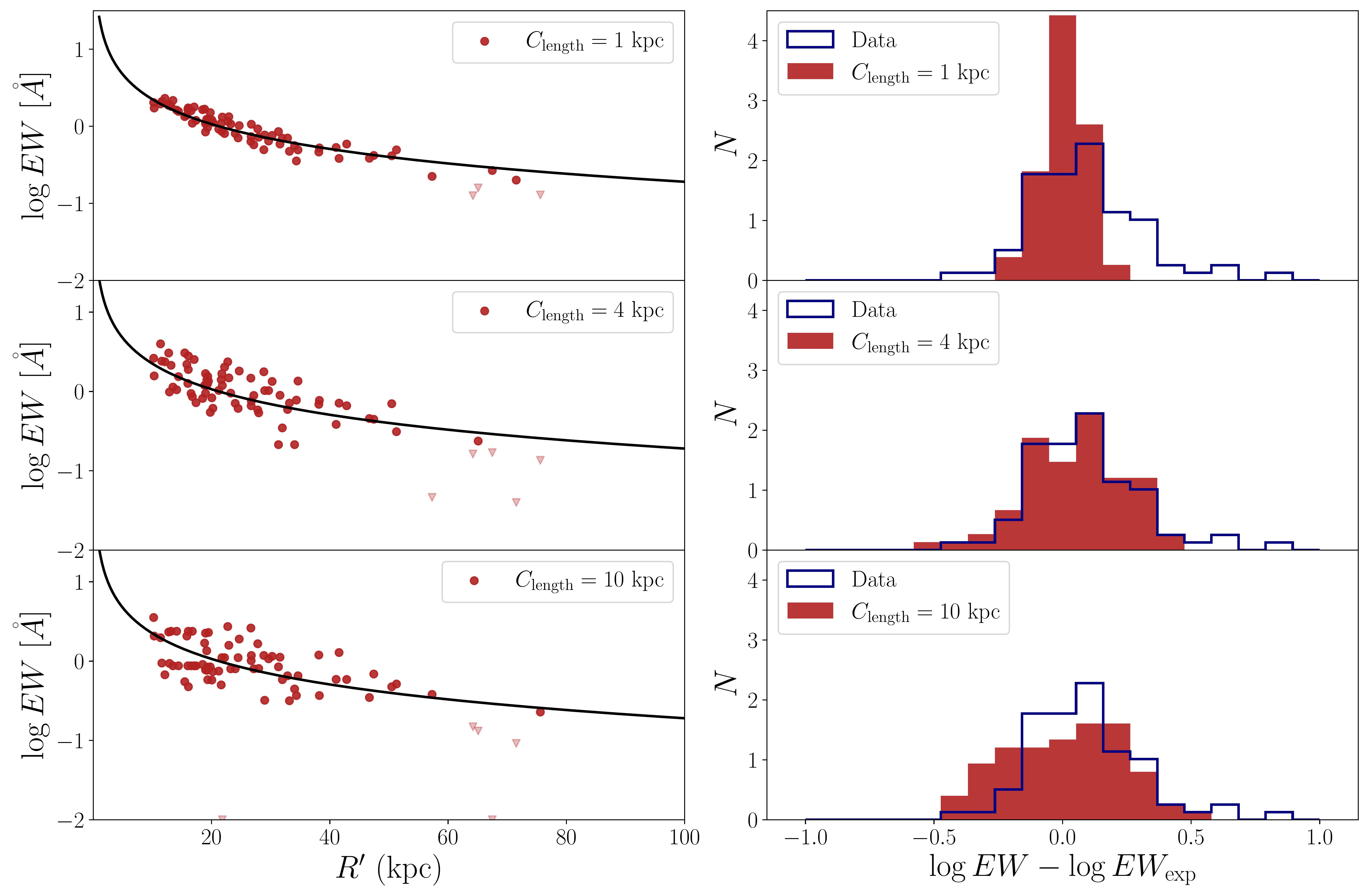}
   \caption{Outputs of the three models presented in Fig.~\ref{fig:EWmaps}, used to infer the likelihood term $\mathcal{L}_{\rm{scatter}}$. From top to bottom, the models have a coherence length of 1, 4 and 10 kpc. Left: model EW distributions as a function of $R'$, with the black curve showing the relation of Eq.~\eqref{eq:wR}. Downward arrows show spaxels whose predicted EWs are below the detection limit of 0.2 \AA\ and are therefore classified as `model non-detections'. Right: distributions of the differences between the EW values and the corresponding values predicted by Eq.~\eqref{eq:wR}, for data (blue) and models (red).}\label{fig:ComparisonRubin}%
    \end{figure*}

The outputs of the method described above are shown in Fig.~\ref{fig:EWmaps}, with their dependence on the gas coherence length. In particular, $C_{\rm{length}}$ is equal to 1, 4 and 10 kpc going from top to bottom. We can see that, even though the general trend is similar (by construction), the maps look quite different between each other, due to the different scale over which the EW is allowed to vary. Note how all maps show `holes', resembling the assumption that the cool CGM is made by clouds whose total covering fraction is lower than one. Both $C_{\rm{length}}$ and covering fractions have an effect on the ARCTOMO \mgii\ detections.

\subsection{Synthetic observations}\label{syntheticObs}
To use the observational constraints outlined in Sect.~\ref{mainObs} one needs to perform synthetic observations. To this aim, we mimic the MUSE observations by creating distributions of spaxels that follow the configuration of our three fields combined, correcting the position of each spaxel vertex for the galaxy mass (see Eq.~\ref{eq:Rcorr}) and choosing a random orientation angle. An example of these distributions is shown in black overlapped to the maps of Fig.~\ref{fig:EWmaps}. We select only those spaxels for which we have detections in the data.

We then extract, from each one of these spaxels, the resulting equivalent width, given by the average of all the EW values (both `systems' and `holes') intercepted by the spaxel, each of them weighted by the fraction of the cell area that overlaps with the spaxel. We exclude all the spaxels within 10 kpc from the center: at these distances Eq.~\eqref{eq:wR} is indeed only an extrapolation and we have seen in Fig.~\ref{fig:EWvsRbins} that our arctomographic data depart from this prediction. This region therefore does not satisfy the assumption that our data are consistent with the results from \cite{huang21}. We have anyway verified that including these data does not affect the results reported in Sect.~\ref{results}. The final EW distribution represents the model output that we can compare directly with the observations, in order to constrain $C_{\rm length}$, as we describe in detail in the following section.
\subsection{Comparison with data and likelihood definition}\label{likelihood}
   \begin{figure*}[!h]
   \includegraphics[clip, trim={0cm 0cm 0cm 0cm}, width=\linewidth]{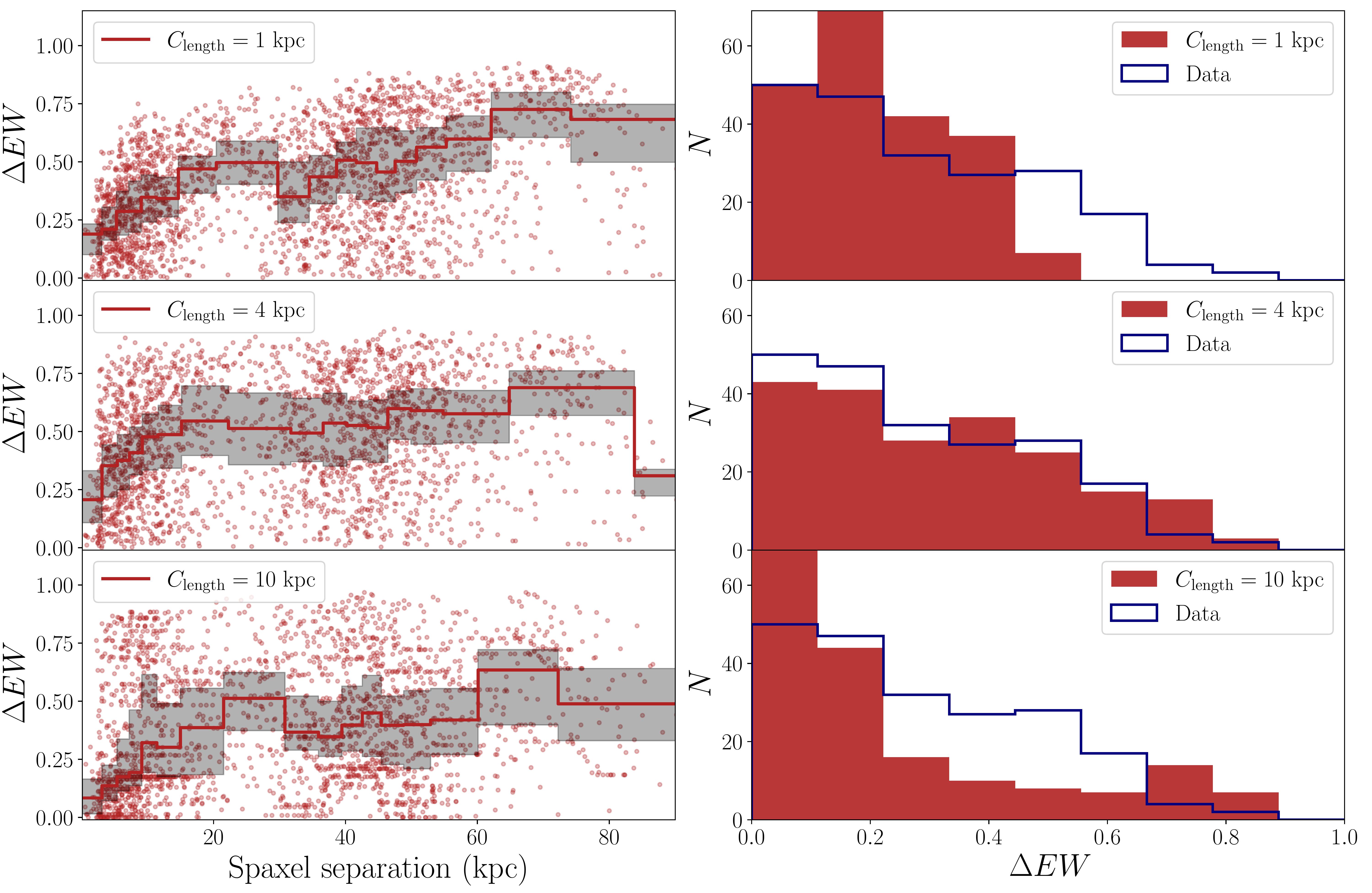}
   \caption{Outputs of the three models presented in Fig.~\ref{fig:EWmaps}, used to infer the likelihood term $\mathcal{L}_{\rm{fractional}}$. From top to bottom, the models have $C_{\rm{length}}= 1, 4, 10$ kpc. Left: distributions of the fractional differences of EW as a function of the spaxel separation (same as right panel of Fig.~\ref{fig:EWvsRbins}, but for the models). Right: distributions for all the pairs with spaxel separations smaller than 5 kpc, for data (blue) and models (red).}
              \label{fig:fracEWComparison}%
    \end{figure*}
The likelihood  
is composed of two different terms, related to the two main observational constraints reported in Sect.~\ref{mainObs}, which we call $\mathcal{L}_{\rm{scatter}}$ and $\mathcal{L}_{\rm{fractional}}$ and that are explained in detail below. The total likelihood is 
given by:
\begin{ceqn}
\begin{equation}\label{eq:likelihood}
\ln{\mathcal{L}_{\rm{tot}}}  = \ln{\mathcal{L}_{\rm{scatter}}} + \ln{\mathcal{L}_{\rm{fractional}}}\ .
\end{equation}
\end{ceqn}
$\mathcal{L}_{\rm{scatter}}$ is used to compare the variation of EW as a function of the impact parameter, focusing in particular on the intrinsic scatter of this distribution with respect to the best fit relation, as this represents a proxy for the gas coherence length. This can be seen in Fig.~\ref{fig:ComparisonRubin}, where we show the results of our synthetic observations 
on the three different model EW maps shown in Fig.~\ref{fig:EWmaps}. 
On the left panels, we can see the EW distributions extracted from our models as a function of $R'$, for a coherence length of 1 (top), 4 (center) and 10 kpc (bottom). The downward triangles depict the `model non-detections' (excluded from our analysis), which we define as equivalent widths lower than the observational detection limit in the arctomographic data, equal to $0.2$ \AA\ (see Table~\ref{table_data}). Note how the predictions of the three models follow the best-fit relation of Eq.~\eqref{eq:wR}, as expected by construction. 
The outputs are however very different from each other in terms of the intrinsic scatter of the distributions, which is much smaller for smaller $C_{\rm{length}}$. Indeed, if the scale of variation of the equivalent width is smaller than the extension of the source, different EW values will be averaged along the source extension, resulting in an overall smaller scatter than if the coherence scale was larger than the source. This was hypothesized and then demonstrated by \cite{rubin18}, where they performed such an analysis for the first time, using galaxies as extended background sources; here 
our extended sources are the MUSE spaxels.

On the right-hand column of Fig.~\ref{fig:ComparisonRubin}, we show 
in red the binned differences ($\log EW - \log EW_{\rm{exp}}$) between our synthetic data and the expectations from Eq.~\eqref{eq:wR} (i.e., the black curve in the left panels), for the same models shown on the corresponding left panels. 
The blue histogram shows the distribution of the same quantity for the observational data. 
Both in models and data we exclude the values with $R'<10$ kpc, for the reasons discussed above.
We can see how the modeled distributions in the right panels of Fig.~\ref{fig:ComparisonRubin} differ from each other, as models with larger coherence lengths yield wider distributions.  
By comparing models and data, we can therefore estimate which coherence length is in a better agreement with the observations.
We finally define the likelihood term $\mathcal{L}_{\rm{scatter}}$ as the p-value obtained by comparing the $\log EW - \log EW_{\rm{exp}}$ distributions of models and data with a Kolmogorov-Smirnov (KS) test\footnote{We also performed an Anderson-Darling test \citep{anderson54}, which produces very similar results.} \citep{massey51}.

$\mathcal{L}_{\rm{fractional}}$ is instead related to the comparison of the fractional differences.  
In Fig.~\ref{fig:fracEWComparison} we show the outputs of our synthetic observations for the three maps shown in Fig.~\ref{fig:EWmaps}, for three different coherence lengths of 1, 4 and 10 kpc. The left panels show the distribution of $\Delta EW$ as a function of the spaxel separation, calculated in the same way as for the data, explained in Sect.~\ref{mainObs} and shown in the right panel of Fig.~\ref{fig:EWvsRbins}. We can see how the shape of the distribution varies when changing $C_{\rm{length}}$. To compare with the data, we focus only on the small spaxel separations. Indeed, while we have seen that our isotropic model is a good representation of our observations, any deviation in the data from this isotropic assumption (see Sect.~\ref{limitations}) will affect more significantly the shape of the $\Delta EW$ distribution at large separations, which will be therefore not consistent with the predictions of our models. By focusing only on small separations, we are more confident that any difference between model and data is entirely due to the cool gas coherence scale.

In detail, we select all the pairs with spaxel separations smaller than 5 kpc and the resulting distributions for the three models are shown in red in the right panels of Fig.~\ref{fig:fracEWComparison}, with overlapped in blue the distribution obtained from the data. We can see how the width of the model distribution increases when increasing the coherence length from 1 to 4 kpc. Indeed, if $C_{\rm{length}}$ is significantly smaller than the length of the spaxel, different EW values will be averaged with each other, with the overall effect of reducing the differences between the EW values of different nearby spaxels, with respect to models where $C_{\rm{length}}$ is more similar to the spaxel size. However, models with $C_{\rm{length}}=10$ kpc have a large amount of low $\Delta EW$, since, if the coherence scale is considerably larger than the length of the spaxels, adjacent spaxels will be probing the same equivalent width values, significantly reducing $\Delta EW$. Therefore, this observational constraint seems to represent a good diagnostic to infer the coherence scale of the cool CGM. 
The likelihood term $\mathcal{L}_{\rm{fractional}}$ is given by the p-value of a KS test performed between the model and data distributions shown in the right panels of Fig.~\ref{fig:fracEWComparison}.

\subsection{Bayesian analysis}\label{Bayes}
Having entirely determined the likelihood of Eq.~\eqref{eq:likelihood}, we can use it to 
constrain $C_{\rm length}$. 
We perform a Bayesian analysis, using the nested sampling method adopting the python package {\sc Dynesty} \citep[][]{skilling04}, over our only free parameter $C_{\rm{length}}$. We utilize the likelihood function $\mathcal{L}_{\rm{tot}}$ defined in Sect.~\ref{likelihood} (Eq.~\ref{eq:likelihood}) and a uniform prior on a range that varies from $0.5$ to $12$ kpc. In fact, just by simply looking at the observed EW maps of Fig.~\ref{fig:EW_maps_orig}, one can already see large EW variations on adjacent spaxels, which implies that $C_{\rm{length}}$ must be of the order of a spaxel linear size (a few kpc, see Table~\ref{table_data}). If $C_{\rm{length}}\ll 1$ kpc or $C_{\rm{length}}\gg 10$ kpc, we would indeed expect a much smoother EW variation across the halo. Therefore, with this prior we are able to explore the full range of possible cool gas coherence scales that might be in agreement with the data. In each call, once the value of $C_{\rm{length}}$ is chosen, to estimate the likelihood we extract a random realization of the model and the data. The latter is obtained by  
randomly drawing the values of the observed EWs from a Gaussian distribution centered on the EW value and with a standard deviation equal to its error \citep[see also][]{rubin18} and by assuming orientation angles for the three fields equal to those used to generate the synthetic observations from the models. 

\section{Results}\label{results}

\subsection{Coherence scale of the cool CGM}\label{ResCl}
Before reporting the results of the Bayesian analysis, we look at how our models compare with the two observational constraints as a function of the assumed coherence length, by analyzing separately the two parts of our likelihood, $\mathcal{L}_{\rm{scatter}}$ and $\mathcal{L}_{\rm{fractional}}$. Indeed, while we use the Bayesian analysis to accurately constrain the range of $C_{\rm{length}}$ that better reproduces the observations, it is important to know whether or not the predictions of our models are effectively in agreement with the data.

\begin{figure}
   \includegraphics[clip, trim={0cm 0cm 0cm 0cm}, width=\linewidth]{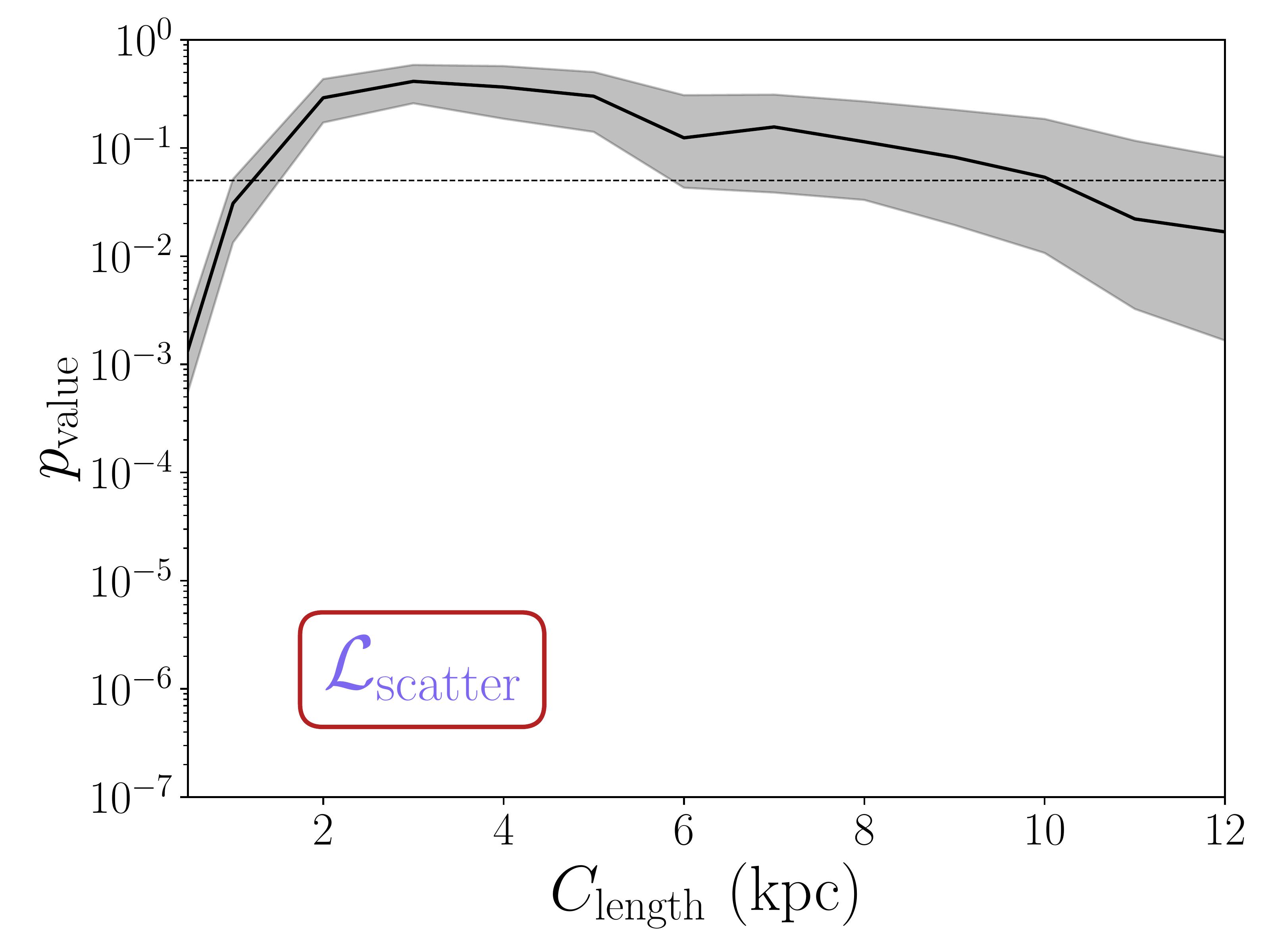}
   \includegraphics[clip, trim={0cm 0cm 0cm 0cm}, width=\linewidth]{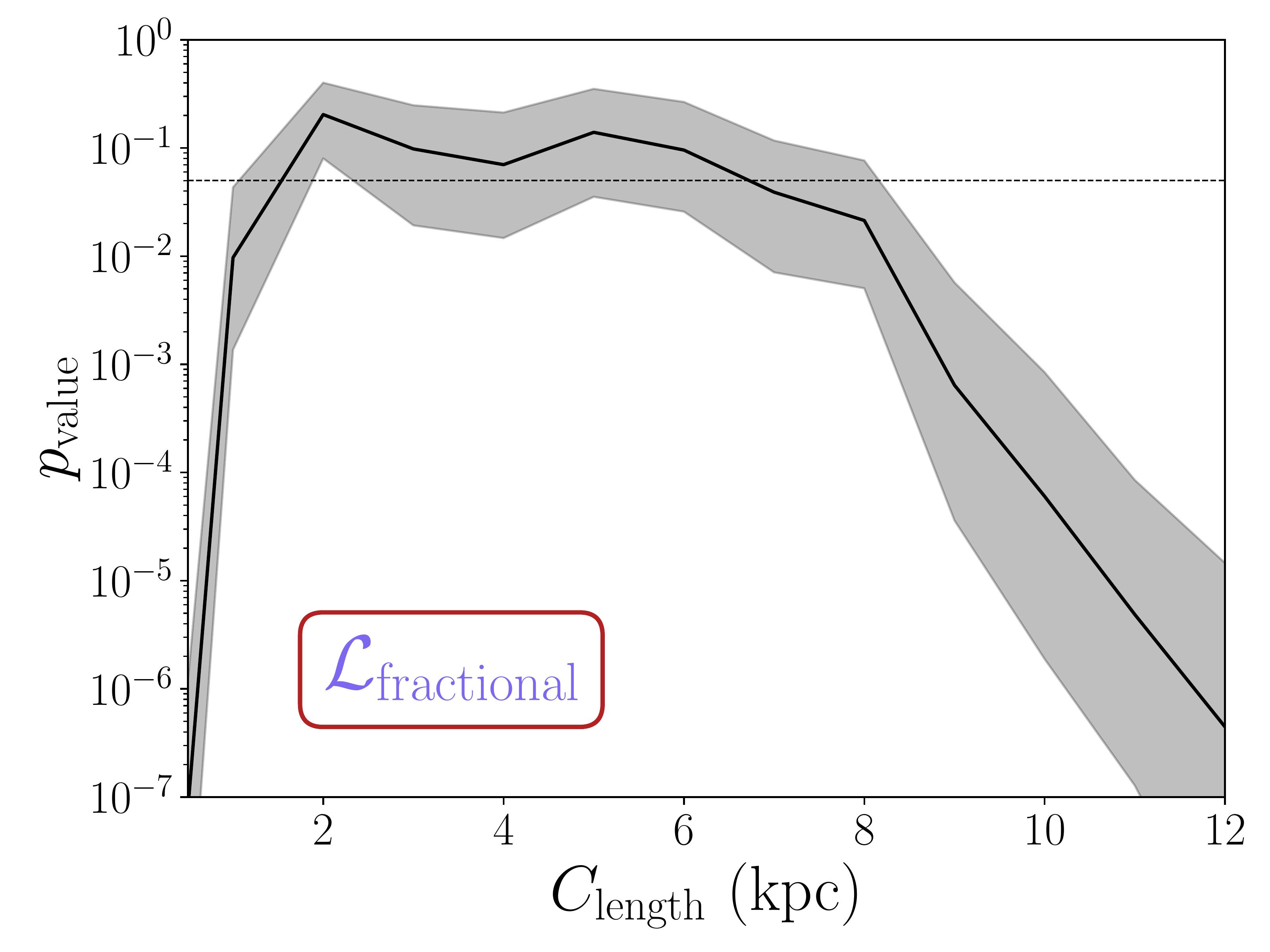}
   \caption{Comparison between models with different coherence lengths and the ARCTOMO data: the two plots show the p-values of the KS test between model and data, as a function of the coherence length adopted in the model. The black curve shows the median value of the 1000 realizations created for each $C_{\rm{length}}$ value (see Sect.~\ref{ResCl}), with the gray band showing the 32nd and 68th percentiles. The horizontal dotted line depicts $p_{\rm{value}}=0.05$, so that models with a probability higher than this threshold can be considered in agreement with the data. Top, comparison of the scatter in the EW-vs-$R'$ relation; bottom, comparison of the EW fractional differences for spaxel separations smaller than 5 kpc. This figure shows that models with $1.5\lesssim C_{\rm{length}}/\rm{kpc}\lesssim7$ are consistent with the data.}
              \label{fig:pvalues}%
    \end{figure}

We explore values of the coherence length varying from 0.5 to 12 kpc and, for each choice of $C_{\rm{length}}$, we create 1000 model and data realizations. 
We finally compare, for each realization, models and data using a KS test, as in Sect.~\ref{likelihood}. The top panel of Fig.~\ref{fig:pvalues} shows the results related to the scatter in the EW-vs-$R'$ relation ($\mathcal{L}_{\rm{scatter}}$) and the bottom panel those related to the distributions of fractional difference of EW ($\mathcal{L}_{\rm{fractional}}$). The black curves show the median values of the KS-test p-value, while the gray bands show the extent of the 32nd and 68th percentiles of the distributions. We consider models with p-values higher than 0.05 (marked by the horizontal dashed line) in agreement with the data.

   \begin{figure}
   \includegraphics[clip, trim={0cm 0cm 0cm 0cm}, width=\linewidth]{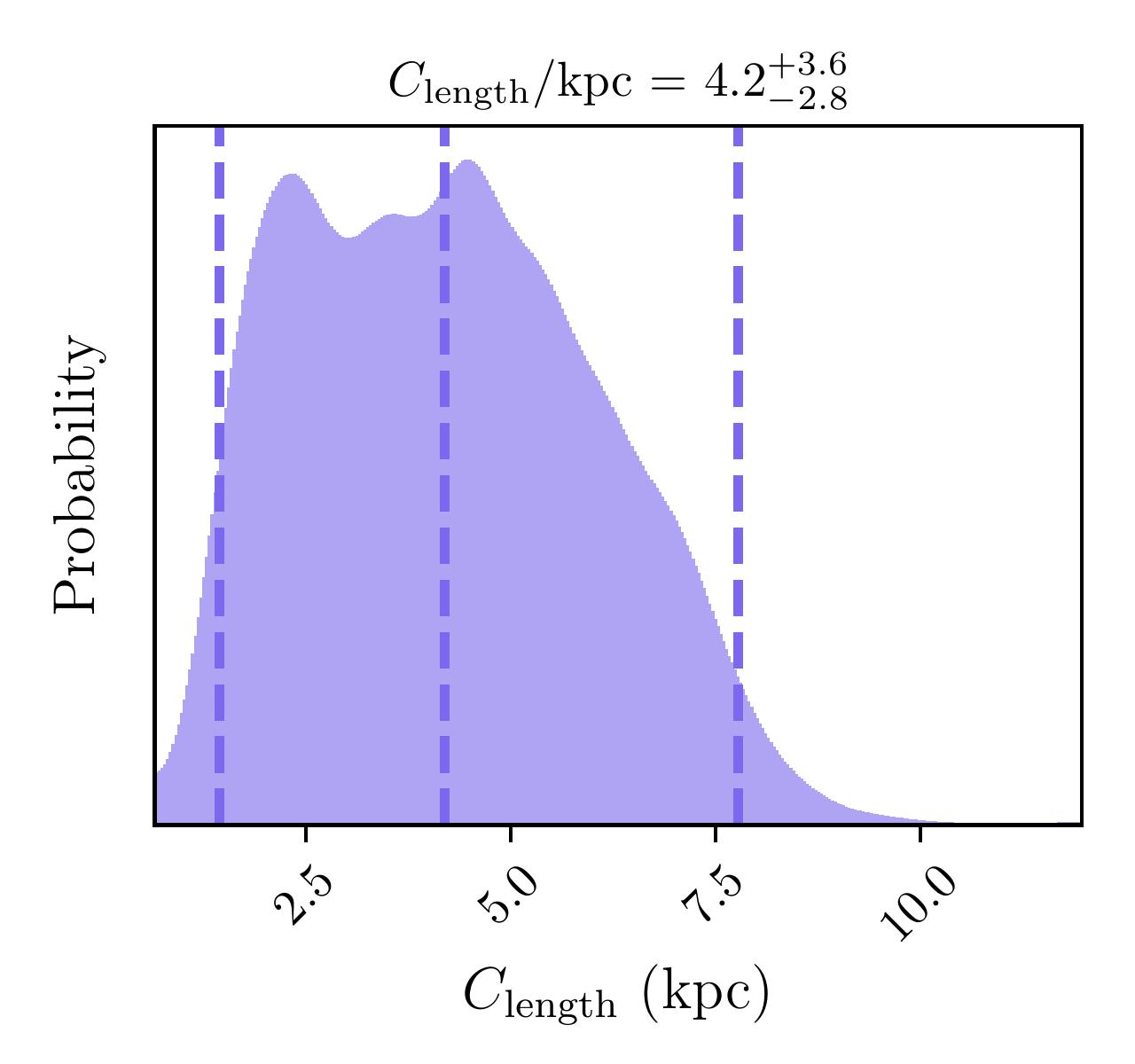}
   \caption{Posterior distribution of the coherence length, as found with our Bayesian analysis. The vertical lines show the 2.5, 50 and 97.5 percentiles of the distribution, corresponding to the median value and its 2$\sigma$ uncertainties, whose values are shown on top. The ARCTOMO data are best reproduced by coherence lengths varying from 1.4 to 7.8 kpc.}
              \label{fig:DynestyResults}%
    \end{figure}

We can see how the consistency with the data depends on the adopted value of $C_{\rm{length}}$. In particular, the constraint given by the scatter in the EW-vs-$R'$ relation (top panel of Fig.~\ref{fig:pvalues}) can be used to put a lower limit on the value of the coherence length, at about 1.3 kpc (where the median p-value is equal to 0.05), in line with the findings of \cite{rubin18}, who used the same constraint to evaluate the coherence scale of the cool CGM. The upper limit on $C_{\rm{length}}$, using this constraint, is at about 10 kpc or more. The comparison with the fractional differences of equivalent width (bottom panel) allows us to put again a lower limit at about 1.5 kpc and a much tighter constraint on the upper limit of the coherence length, at about 7 kpc, and represents therefore a more stringent constraint for $C_{\rm{length}}$. Together, the two constraints define a range of models with $1.5\lesssim C_{\rm{length}}/\rm{kpc}\lesssim7$ that are in agreement with the ARCTOMO data. 

We move then to the more quantitative results of the Bayesian analysis, where we adopt the full likelihood described by Eq.~\eqref{eq:likelihood}.
The main result of this work is shown in Fig.~\ref{fig:DynestyResults}, where we report the posterior distribution of the coherence length, the free parameter in our analysis. The vertical dashed lines show the value of the median of the distribution, with the 2$\sigma$ uncertainties (0.025 and 0.975 quantiles, see also Table~\ref{tab:testsCL}). We can see that  
$C_{\rm length}$ is well defined between 1.4 and 7.8 kpc, 
with a median value equal to 4.2 kpc. The 2$\sigma$ limits of the posterior distribution capture the whole range of models that are in agreement with the ARCTOMO data, as shown in the two panels of Fig.~\ref{fig:pvalues}. 
These are the tightest bounds on $C_{\rm length}$ so far. 
We stress that this is possible thanks to the unique nature of the ARCTOMO data, which allow us to have multiple and extended sources (the MUSE spaxels) around individual galaxies. 

\subsection{How universal is our result?}

 Given the complexity of the gaseous halos of galaxies, one could expect the coherence length of the cool CGM to depend for example on its distance from the central galaxy or to vary across different galaxy halos.  
In the following sections we investigate more in detail these aspects, by analyzing sub-samples of the original set of data that was used to obtain the results reported above. 

\subsubsection{Variation of $C_{\rm{length}}$ with impact parameter}\label{ResClvary}
   \begin{figure*}
   \centering
   \includegraphics[clip, trim={0cm 0cm 0cm 0cm}, width=0.33\linewidth]{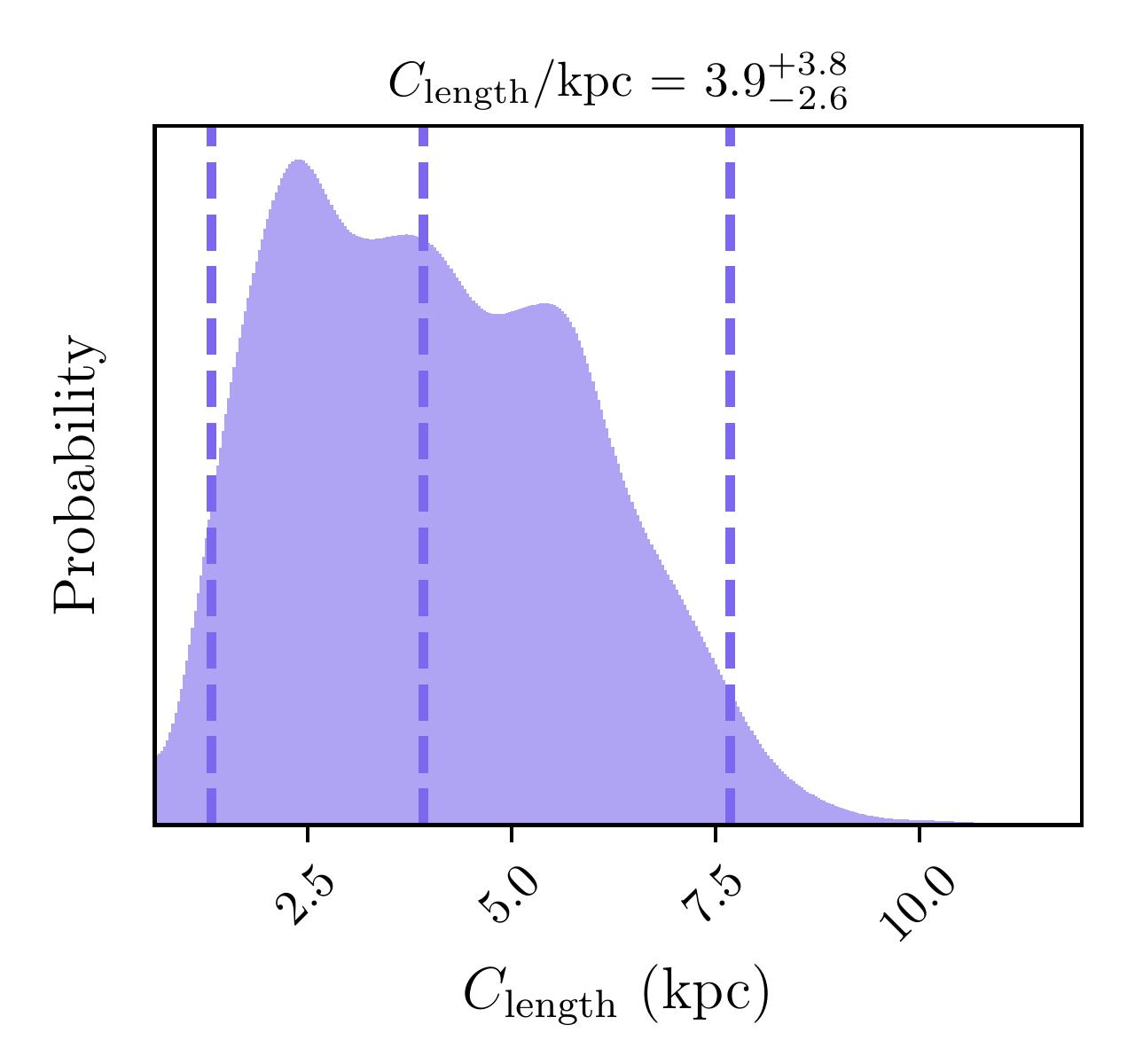}
   \includegraphics[clip, trim={0cm 0cm 0cm 0cm}, width=0.33\linewidth]{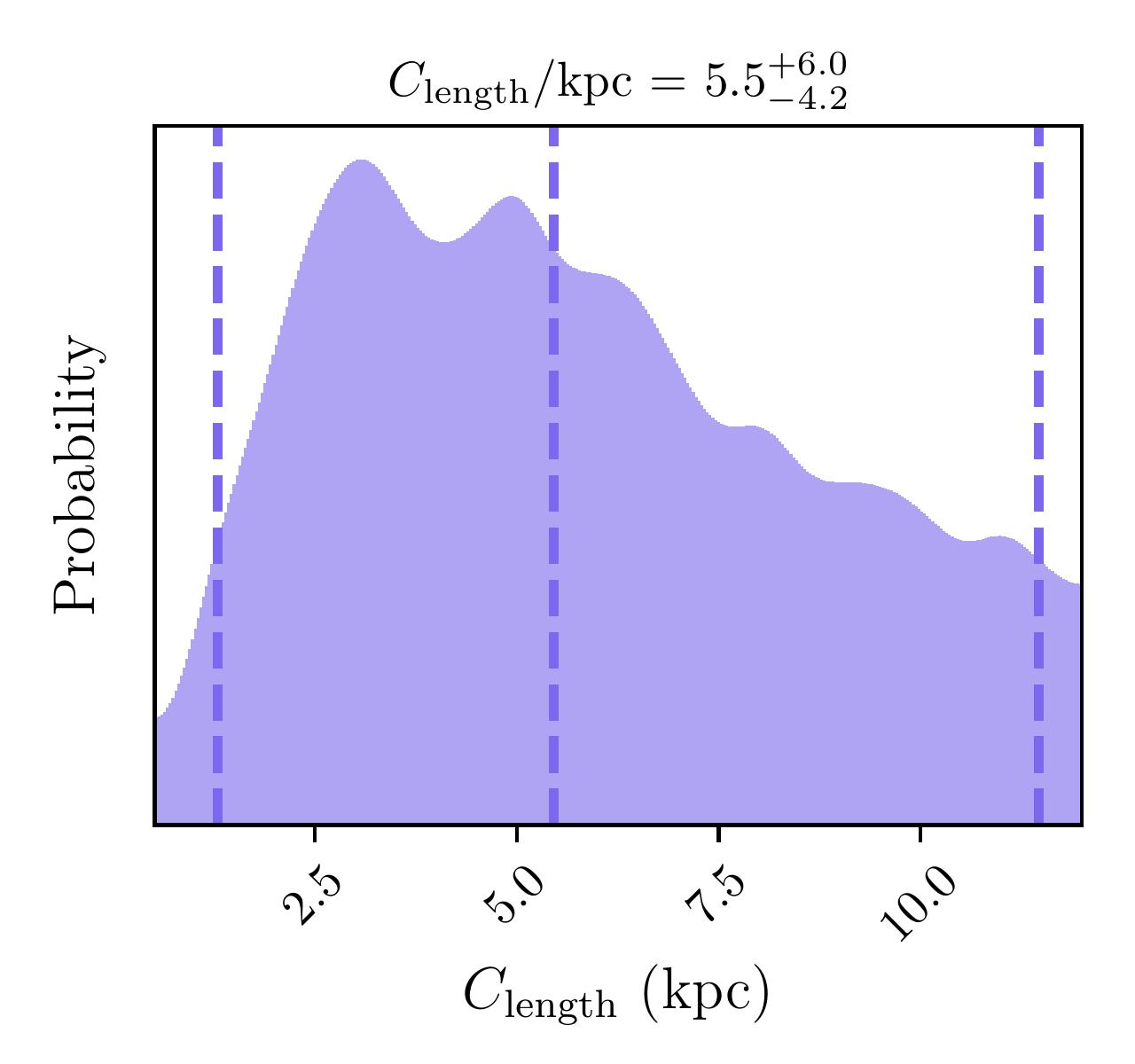}
   \includegraphics[clip, trim={0cm 0cm 0cm 0cm}, width=0.33\linewidth]{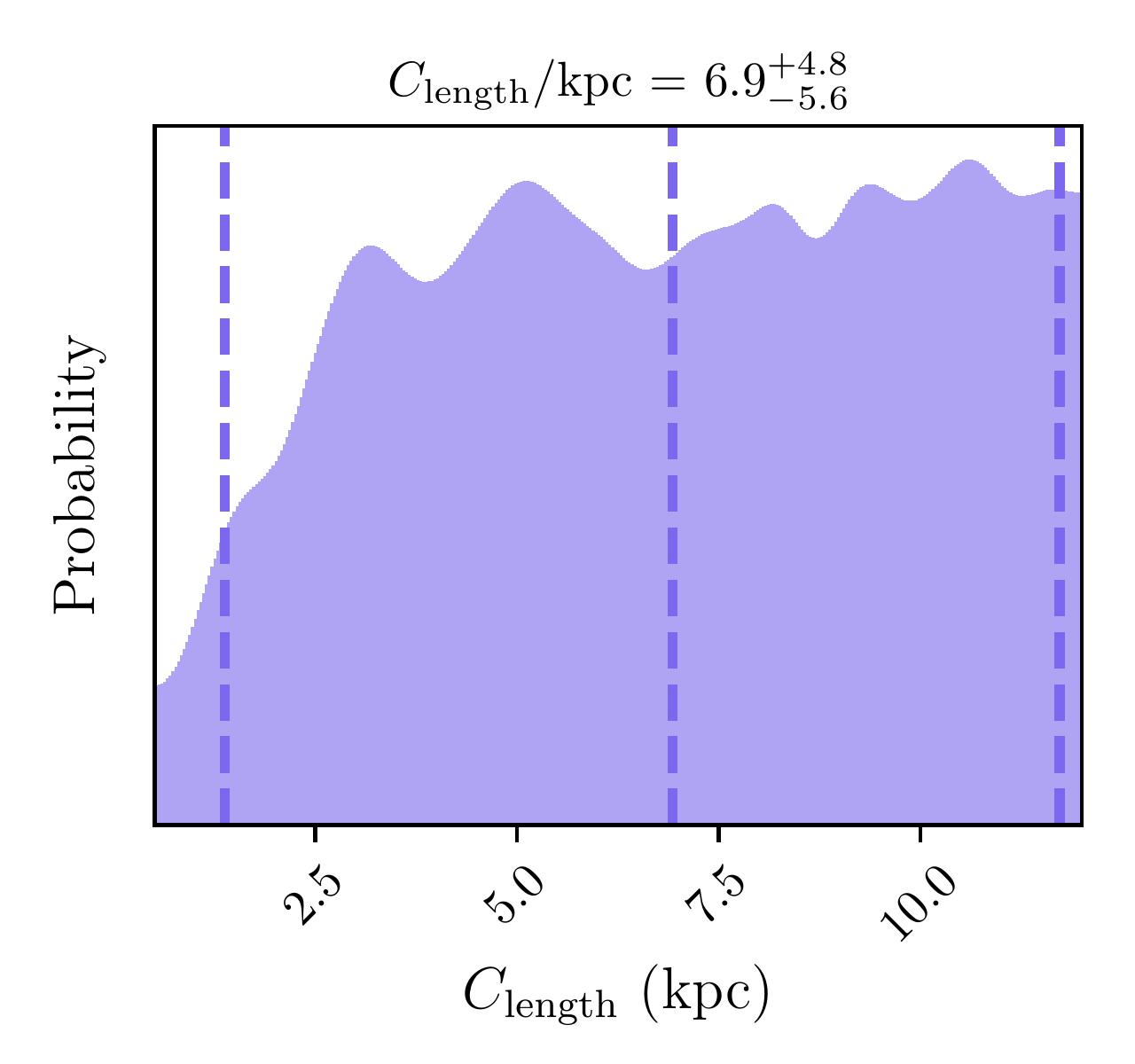}
   \caption{Same as Fig.~\ref{fig:DynestyResults}, but for the three sub-samples obtained by selecting the spaxels with $R'<30$ kpc (left), $30<R'/\rm{kpc}<50$ (center) and $R'>50$ kpc (right).}
              \label{fig:dynRcorr}%
    \end{figure*}
   \begin{figure*}
   \includegraphics[clip, trim={0cm 0cm 0cm 0cm}, width=0.33\linewidth]{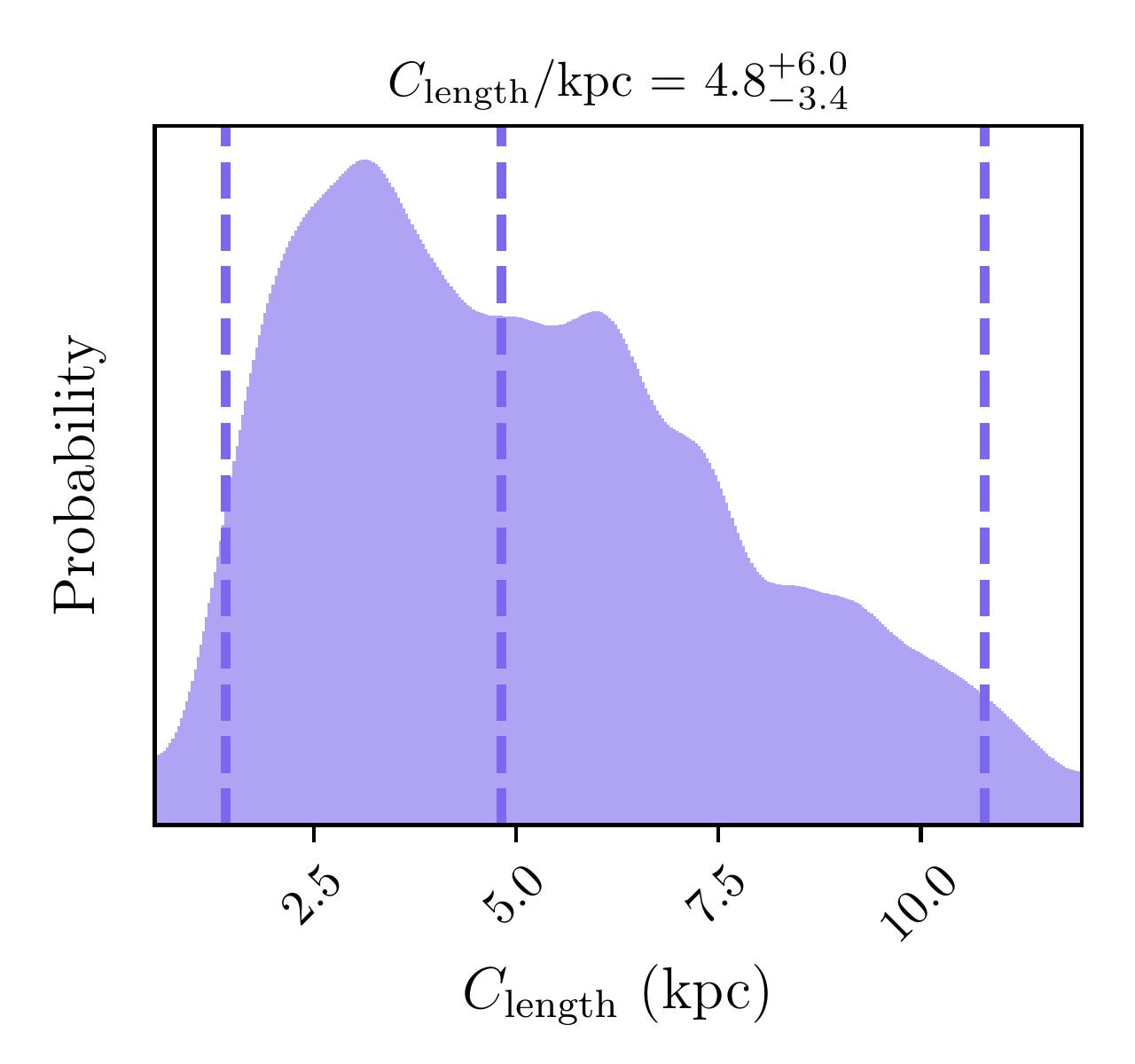}
   \includegraphics[clip, trim={0cm 0cm 0cm 0cm}, width=0.33\linewidth]{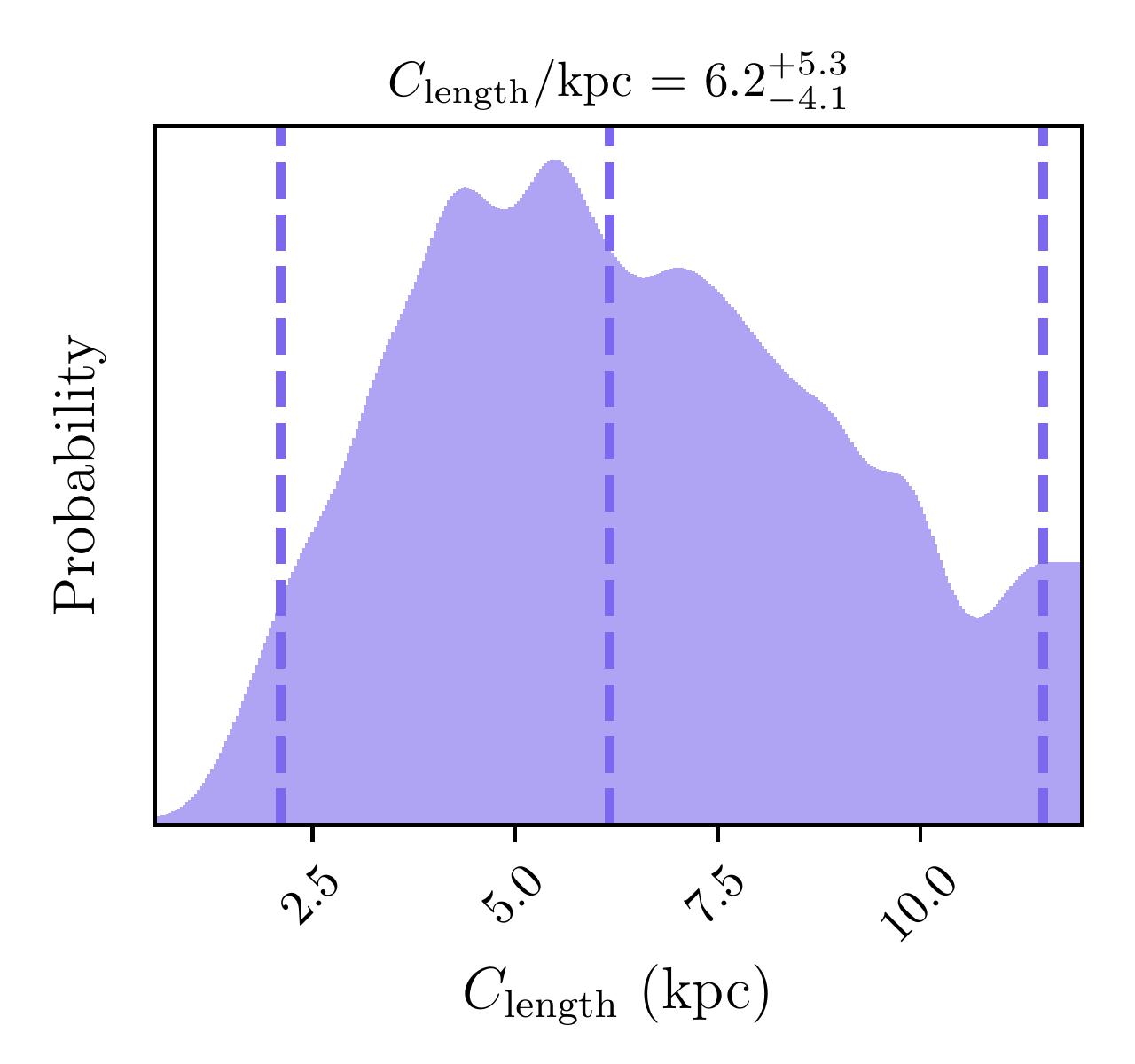}
   \includegraphics[clip, trim={0cm 0cm 0cm 0cm}, width=0.33\linewidth]{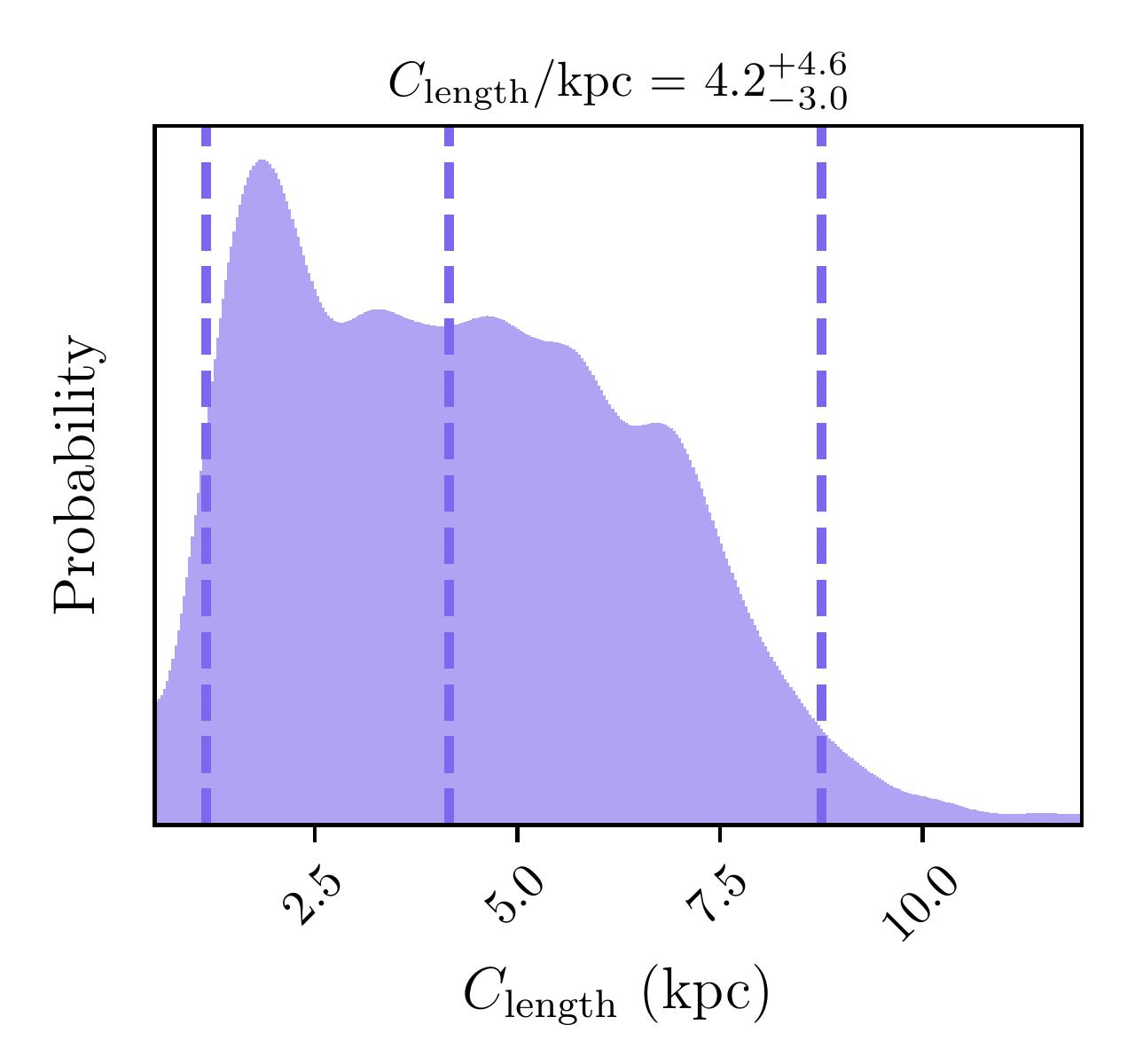}
   \caption{Same as Fig.~\ref{fig:DynestyResults}, but selecting separately the spaxels of the three MUSE fields: \pks\ (left), \rcs\ (center) and \sgas\ (right).}
              \label{fig:dynGalaxies}%
    \end{figure*}

We have seen that at least some of the properties of the cool CGM, like EW and covering fraction 
\citep[e.g.,][]{huang21}, seem to be related to its projected distance from the galaxy host (or impact parameter). Even though it is important to keep in mind that these are all projected quantities, such trends might be related to the origin and dynamics of the cool CGM and to its connection with the central galaxy. Using our dataset, we investigate whether a similar dependency is present also for the gas coherence length.

We divide our original sample in 
three bins, for spaxels with $R'<30$ kpc, $30<R'/\rm{kpc}<50$ and $R'>50$ kpc. For each sub-sample we then perform the same Bayesian analysis performed on the whole sample. The results of this procedure are shown in Fig.~\ref{fig:dynRcorr} and in Table~\ref{tab:testsCL}. We find a slight trend of increasing coherence length at larger projected distances from the central galaxies, with the median values of the posterior distributions going from 3.9 kpc in the innermost bin to almost 7 kpc in the outermost one. However, it is hard to say whether this trend is physical or simply due to the low-number statistics. The three posterior distributions share indeed the same value of the 2.5 percentile (1.3 kpc) and the main difference is in the distribution width. The large `tails' at large $C_{\rm{length}}$ in the middle and right panels are likely due to the fact that most of the spaxels in our sample are within 30 kpc from the galaxy (47 detections for $R'<30$ kpc, 21 detections for $30<R'/\rm{kpc}<50$ and 7 detections for $R'>50$ kpc) and therefore the constraints on $C_{\rm{length}}$ at large $R'$ are very loose\footnote{Note also that, extending the prior on $C_{\rm{length}}$, the tails shown in the middle and right panels of Fig.~\ref{fig:dynRcorr} would be even more extended and the median and 97.5 percentile values would increase. This is most likely not due to a physical property of the CGM, but to the small number of spaxels present at large $R'$.}. 
 
Given the considerations above, we argue that, if any, the dependence of the coherence length with the impact parameter is only minor. This could imply that the cool CGM is originated by similar physical processes across the galactic halo, which tend to create similar gas structures independently of the distance from the central galaxies. More data will be needed to study this in a greater detail. 

\subsubsection{Variation of $C_{\rm{length}}$ between single galaxies}\label{galvary}

We perform the same analysis described in Sect.~\ref{Bayes}, but selecting the three halos separately, selecting each time only the spaxels from one of the three MUSE fields. The results of these tests are shown in Fig.~\ref{fig:dynGalaxies} and Table~\ref{tab:testsCL}. We can note how the posterior distributions of the gas coherence length are slightly different in the three cases, indicating that this property can vary across different galaxy halos. While the distributions of the \pks\ and the \sgas\ fields are very similar, with the main difference being the smaller width of the latter (due to the larger number of detections in \sgas), the posterior distribution for the \rcs\ field seems to point to a larger gas coherence length in the cool CGM of this galaxy with respect to the other two. 
This difference seems however to not be significant, given that in all cases the posterior distributions are relatively extended and the three coherence lengths are consistent with each other when considering the uncertainties. The slight difference in the $C_{\rm{length}}$ of \rcs\ might be due to the fact that the spaxels in this field are on average at larger impact parameters with respect to the other two fields, possibly reinforcing the idea of a connection between $C_{\rm{length}}$ and the impact parameter. A larger number of spaxels for each halo would be needed to more accurately constrain the coherence length of the cool CGM of single galaxies.

It is particularly important to note that the three median values of the distributions are all within the range of coherence lengths that we find when combining the three fields together (Fig.~\ref{fig:DynestyResults}). 
This suggests 
that the range of coherence lengths between about 1.5 and 8 kpc that we found in Sect.~\ref{ResCl} is not driven by one particular galaxy, but likely represents a common property of the cool CGM of these three objects (as we initially assumed). This result seems to confirm that similar galaxies are surrounded by cool CGM with a similar structure.  
A larger sample of galaxies will be needed to confirm this finding.
{ 
\renewcommand{\arraystretch}{1.5}
 \begin{table}[htbp]
\centering
\caption{Results of the Bayesian analysis.}\label{tab:testsCL}
\begin{tabular}{*{4}{c}}
\hline  
\hline
 TEST & 2.5 perc. & Median & 97.5 perc. \\
  & (kpc) & (kpc) & (kpc) \\
\hline 
Full sample&1.4 & 4.2 & 7.8 \\
$R'/\rm{kpc}<30$&1.3 & 3.9 & 7.7 \\
$30<R'/\rm{kpc}<50$&1.3&5.5&11.5\\
$R'/\rm{kpc}>50$&1.3&6.9&11.7\\
\pks &1.4&4.8&10.8\\
\rcs &2.1&6.2&11.5\\
\sgas &1.2&4.2&8.8\\
Free $\sigma$&1.3&4.5&8\\
$4\times4$ binning&1.8&5.7&10.6\\
\hline
\end{tabular}
\vspace{0.1cm}
\flushleft
 Notes: 2.5 percentile, median and 97.5 percentile of the posterior distribution of $C_{\rm{length}}$, for the fiducial analysis of the whole sample (Sect.~\ref{ResCl}), the tests performed on different bins of projected distance (Sect.~\ref{ResClvary}) and on the three individual galaxy halos (Sect.~\ref{galvary}), and the tests where we introduced the intrinsic scatter $\sigma$ as an additional free parameter and where we used a $4\times4$ binning in the data (see Sect.~\ref{limitations}).
   \end{table}
   }
\section{Discussion}\label{discussion}

\subsection{What is the coherence scale?}\label{CL_discussion}
We have constrained the typical scale over which the absorption strength of the cool CGM around three galaxies at $z\lesssim1$ does not vary. From an observational point of view, a coherence scale, or length, of a few kpc means that when observing the cool CGM around a single galaxy with multiple probes, we do not expect to observe strong variations in \mgii\ EW for (projected) separations smaller than this scale, while we expect little correlation for larger separations. This can be seen in the right-hand-side of Fig.~\ref{fig:EWvsRbins} and has been found previously by other studies (see Sect.~\ref{previousWorks}). The fact that we find an entire range of coherence lengths that seem to well reproduce our data ($1.4\lesssim C_{\rm{length}}/\rm{kpc}\lesssim 7.8$), means that likely the exact value of this scale may vary across the halo, although we found that there is little (if any) dependence of $C_{\rm{length}}$ with the projected distance from the central galaxy. To better understand the variation of the coherence length across the halos of galaxies, more data will be needed.

What is the physical nature of this scale? As already discussed extensively in \cite{rubin18}, a variation in EW may be due not only to differences in the number of clouds intercepted by the line of sight, but also to differences in the clouds' kinematics (and partly also in gas metallicity). 
Higher spectral resolution observations \citep[e.g.,][]{lopez99,rauch02,zahedy16,krogager18} and/or more sophisticated 3D modeling (see for example \citealt{li23alpaca} and Sect.~\ref{previousWorks}) are needed to distinguish between different kinematic components along the line of sight and to better understand what property is driving the change in EW. 
The coherence length that we derived using our empirical models of EW is instead a projected quantity and does not allow us to make conclusions on the physical properties of the individual CGM clouds. We however speculate that, regardless of the exact property that drives the change in EW, the coherence scale should represent the size of the cool gas structures over which such properties change. Each one of such structures could be composed by a relatively large cloud with a size comparable to $C_{\rm{length}}$, or by multiple smaller clouds whose properties are related to each other and that cluster over a coherence scale. This length is a direct observable that can be tested on upcoming observational data and on theoretical models (see Sect.~\ref{theory}) and it is independent of any physical assumption on the cool CGM.

\subsection{Comparison with previous works}\label{previousWorks}
The main point of comparison for our study is the work of \cite{rubin18}, on which the idea for the present exploration was based.  
In a similar fashion to what we have later developed in this paper, they used a fiducial model for the cool CGM based on the quasar surveys of \cite{chen10new} and \cite{werk13} and they compared the predictions of such model with their data (a sample of 27 galaxy-galaxy pairs), to estimate the gas coherence length. The main constrain was given by the comparison of the scatter in the EW-vs-$R'$ distribution of models and data. This coincides with the first part of the likelihood that we used in this work ($\mathcal{L}_{\rm{scatter}}$).

In our analysis we applied two main refinements to the models of \cite{rubin18}: i) we created  2-dimensional distributions of absorbers across the halo, also considering the covering fraction as a function of the projected distance from the center, while \cite{rubin18} explored `slices’ of the CGM at fixed impact parameters, with a unity covering fraction; ii) we used the actual distribution of MUSE spaxels and their shape to extract the synthetic observations, while \cite{rubin18} used more idealized distributions of circular beams to resemble the spatial extension of their background galaxies. 

Regardless of the above differences, and of the use of a model based on a different survey \citep{huang21}, our results are in perfect agreement with those of \cite{rubin18}, who estimated a lower limit for the coherence length of $\sim1.9$ kpc. Our estimate ($1.4 <C_{\rm{length}}/\rm{kpc} <7.8$) is consistent with this value and especially the plot in the top panel of Fig.~\ref{fig:pvalues} shows that, using the same observational constraint of \cite{rubin18}, we would obtain a very similar result, with very loose constraints on the coherence length upper limit. By using the additional observable given by the equivalent width differences across the spaxel pairs (which was evidently not possible in \cite{rubin18}, given the different observational setup) we were also able to put a firm upper limit on the coherence length of the gas, at about 8 kpc. The similarity between our results and those of \cite{rubin18}, who used around 30 absorbing galaxies for their analysis, further points to the conclusion that the cool CGM of galaxies with similar properties has a similar 'universal' structure (see Sect.~\ref{galvary}).

Various other observational studies have tried to estimate the typical scale of the cool circumgalactic medium, either directly through the use of multiply lensed quasars, or indirectly, using photo-ionization models. Here, we summarize some of the main results of these works and we refer to \cite{rubin18} for a more complete and extensive review (see their Section 3.2). The use of lensed quasars to spectroscopically study the CGM has started more than 40 years ago \citep[e.g.,][]{young81,smette95} and has been extensively adopted in the last decades to investigate the structure of this cool gas \citep[e.g.,][]{rauch99,lopez99,rauch02,ellison04,lopez05,chen14,zahedy16,rubin18lensedQSO,okoshi21}. 
It is not trivial to compare our findings with those of these works, as different studies use different gas properties, like column densities of various ions and/or the gas velocity structure \citep[][]{chen14} to derive a coherence scale of the CGM.  
Even when looking at \mgii\ absorption, the coherence scale is often based only on the number of coincidences and anti-coincidences (detection or non-detection of the absorption) in the QSO lines of sight \citep[e.g.,][]{ellison04}, a method that has been pointed out to have  limitations~\citep{martin10}. Here instead, besides taking full advantage of the arc-tomography spatial sampling, we also use  the variations of the absorption strength across the halo. The general picture that emerges, putting together the results from previous studies, is anyway that weak \mgii\ absorbers (EW $\lesssim 0.4$ \AA) tend to exhibit a smaller coherence scale (less than 1 kpc), with respect to stronger systems (like those present in the ARCTOMO data), which show a coherence length of a few kpc or larger, in agreement with our findings. 

Additional constraints on the structure of the cool CGM have been derived by using photo-ionization models \citep[e.g., CLOUDY][]{ferland98,ferland13,ferland17}, which allow to estimate the thickness of the absorbers\footnote{Note that these models probe the absorber size along the line of sight, while in this work we evaluated the spatial coherence length in the transversal direction (even though these two lengths are not necessarily different from each other).}. Overall, cloud sizes lower than 1 kpc are found at redshifts $z\gtrsim2$ \citep[e.g.,][]{crighton15}, while at $z\lesssim1$ the scales seem in general to be larger, of the order of several up to even hundreds of kpc (e.g., \citealt{werk14,keeney17}, but see \citealt{stern16}, who assuming that the cool gas spans a variety of densities, found that the cool clouds traced by \mgii\ in the COS-Halos sample have sizes of only a few tens of pc). More recently, \cite{zahedy21} applied photo-ionization models to four Lyman-limit systems at redshift lower than 1, taken from the CUBS survey and found that the clouds have thicknesses that can vary from 0.01 to 10 kpc, in agreement with our findings (although their mode is around 0.1 kpc).  It is however important to note that, due to the various assumptions of ionization models, as for example the strength and shape of the adopted UV background \citep[see][]{acharya22,gibson22}, the cloud sizes inferred from these works remain quite uncertain. Our estimate of the coherence length is to date one of the most accurate constraints of the structure of the cool CGM around galaxies at $z\lesssim1$.

In the last years, also more complex and physically motivated models have been used to interpret observations of the cool CGM and to infer, among other properties, the size of the single clouds. \cite{faerman23} built a model of the cool gas assuming pressure equilibrium between the $T\sim10^4$ K medium and a hot corona \citep[see][]{faerman20} and allowing for non-thermal pressure support. They then compared their models with the COS-Halos data \citep[e.g.,][]{werk13} and found that the observed column densities of neutral hydrogen and low and intermediate ions (among which \mgii) are best reproduced by filling factors of the cool gas of the order of 1\%. This, together with the observed number of kinematic components (or clouds) along the line of sight, provides an upper limit for the cloud sizes of $R_{\rm{cl}}\lesssim 0.5$ kpc. This finding is not in  disagreement with our results, since, as discussed in Sect.~\ref{CL_discussion}, we are not modeling single clouds, and each absorbing system in our maps could be made of multiple clouds, which could therefore be smaller than our estimates for the coherence scale. As also mentioned in \cite{faerman23}, multiple clouds could be segregated in larger complexes and structures, consistently with our work and with complementary implications from high-resolution data of the high-velocity clouds in the halo of the Milky Way \citep[e.g.,][]{tripp22}. Finally, using dynamical semi-analytical models of infalling clouds accreted from the intergalactic medium, \cite{afruni22} have reproduced the observed covering fraction, silicon column density and kinematics of the cool absorbers detected by the AMIGA project \citep{lehner20p} in the halo of the Andromeda galaxy. In their best-fit models, typical cool clouds have radii of about 5 kpc at the virial radius and of less than 1 kpc at distances of a few tens of kpc from the central galaxy. Even though we do not have in our empirical models any indication of the intrinsic location of the single absorbing clouds, our coherence length going from about 1.5 to 8 kpc seems to be in perfect agreement with the results of these models. 

\subsection{Implications for hydrodynamical simulations of the CGM}\label{theory}
While the coherence scale that we estimated in this work does not directly reflect the physical 3D distribution and the size of the single clouds, it gives an important constraint on the typical scales over which the cool gas is expected to cluster in separate structures (see Sect.~\ref{CL_discussion}). Our finding represents therefore a very useful observable to compare to current (and future) theoretical models of the circumgalactic medium.

From the point of view of high-resolution hydrodynamical simulations, single clouds can have sizes smaller than a pc, due to processes like fragmentation, or shattering, that might lead to small cloudlets with lengths similar to the gas cooling length \citep[see][]{mccourt18}.  In the hypothesis that these small cloudlets are long-lived, one possible picture is that the cool CGM is composed by a fog of these cloudlets that permeate the whole galaxy halo: this has been proposed especially for the cool CGM at redshifts $z\gtrsim2$, to explain a small filling factor and at the same time a covering fraction close to unity seen in observations \citep[see references in][]{mccourt18}. However, to fully understand the structure of the cool gas one has to consider many other physical effects: small droplets of gas could quickly evaporate into the hot gas that surrounds them, due to thermal conduction \citep{armillotta17,afruni23} or they can coagulate to form larger structures \citep[e.g.,][]{gronke20mist,gronke23}, which could later even grow due to the induced condensation of the hot gas \citep[e.g.,][]{marinacci10,gronke18,kooij21}. The final picture emerging from this type of simulations is then that the cool CGM is likely composed of a spectrum of structures, going from a fog of cloudlets to more `monolithic’ clouds and the predominance of one or the other depends on the physical properties of the gas \citep{gronke20mist}. Our results suggest that, at least in the galaxies probed in this work, the cool CGM presents substantial differences (in \mgii\ EW) on scales between 1.4 and 7.8 kpc. We speculate that this finding can be used by hydrodynamical simulations to put upper limits on the larger scales of the cascade of structures that is believed to form the cool gas around galaxies.

Given their pc-scale resolution, the studies mentioned above can only be focused on small parts of the galactic halo. To have a more extended view, one needs to resort to cosmological and zoom-in simulations, which have been extensively used in the last decade to study the cool CGM \citep[see][for a recent review]{giguere23}. These studies can be very useful to have a full picture of the distribution of the CGM across the halos of galaxies, 
but have necessarily much lower resolutions with respect to the idealized simulations. Even in the best zoom-in cases \citep[e.g.,][]{peeples19} the resolutions can reach a maximum of about 500 pc, which are not sufficient to accurately predict the structure of the cool gas. \cite{vandevoort19} have indeed shown that increasing the resolution in this type of simulations drastically changes the distribution and the covering fraction of the cool gas, indicating no sign of convergence. Therefore, while a considerable effort has been done to compare the predictions of this type of simulations with observations of the cool CGM \citep[see for example][]{appleby21,defilippis21}, the results are still necessarily uncertain. With this caveat in mind, it is worth mentioning that recently cool clouds with sizes of a few kpc, possibly in agreement with our findings, have started to be identified even in cosmological simulations like TNG-50 \citep{nelson20,ramesh23}. A CGM coherence scale could be directly extracted from these simulations and compared with the findings of this study.

Future simulations and in particular works that aim to link small and large-scale simulations \citep[e.g.,][]{huang22,fielding22,weinberger23} will hold the key for a more accurate comparison with the available observational data. Our result regarding the coherence length is one of the fundamental observational constraints that such models should aim to reproduce. 

\subsection{Limitations and assumptions of this study}\label{limitations}
An important assumption we made in this study is that the gas is distributed isotropically across the halo, so that the average strength of the absorption is the same along rings centered on the host galaxy. This is likely a simplistic assumption.
Previous works \citep[e.g.,][]{schroetter19} have found a segregation of absorbers along the galaxy minor and major axis, hinting at the cool CGM being described by a wind plus disk model. In particular, \cite{ferfig22} found a dependence of the EW strength with the azimuthal angle for one of the three MUSE fields analyzed in this work, \sgas, even though the effect dilutes 
when considering the three fields together, as we have seen in Sect.~\ref{mainObs}. Moreover, the presence of satellites in the halo of the main absorbing galaxy might affect the cool gas distribution: part of the absorbing cool gas might indeed come from stripping from these satellite galaxies, as extensively observed in the Local Universe \citep[e.g.,][]{poggianti17,putman21}, thus breaking our assumption of isotropy.

Nevertheless, by considering in our analysis only the EW fractional differences at small spaxel separations (see Sect.~\ref{likelihood}), we are minimizing the effects due to possible anisotropic distributions, 
which occur at larger scales. We have verified that using in our analysis also larger spaxel separations does not significantly affect our main finding on the coherence length range, but it reduces, for any given choice of $C_{\rm{length}}$, the agreement between models and data, indicating a slight level of anisotropy in our data. 
We therefore acknowledge the importance of exploring more in detail also anisotropic models, but we leave it for future studies, as it is outside the scope of this work. 
In future works we will also address more realistic shapes for the gas absorbers, as the squared approximation adopted here is clearly simplistic. However, a more accurate absorber morphology (which is also essentially unknown to date) will likely not affect our findings on the gas coherence scale.

   \begin{figure}
   \includegraphics[clip, trim={0cm 0cm 0cm 0cm}, width=\linewidth]{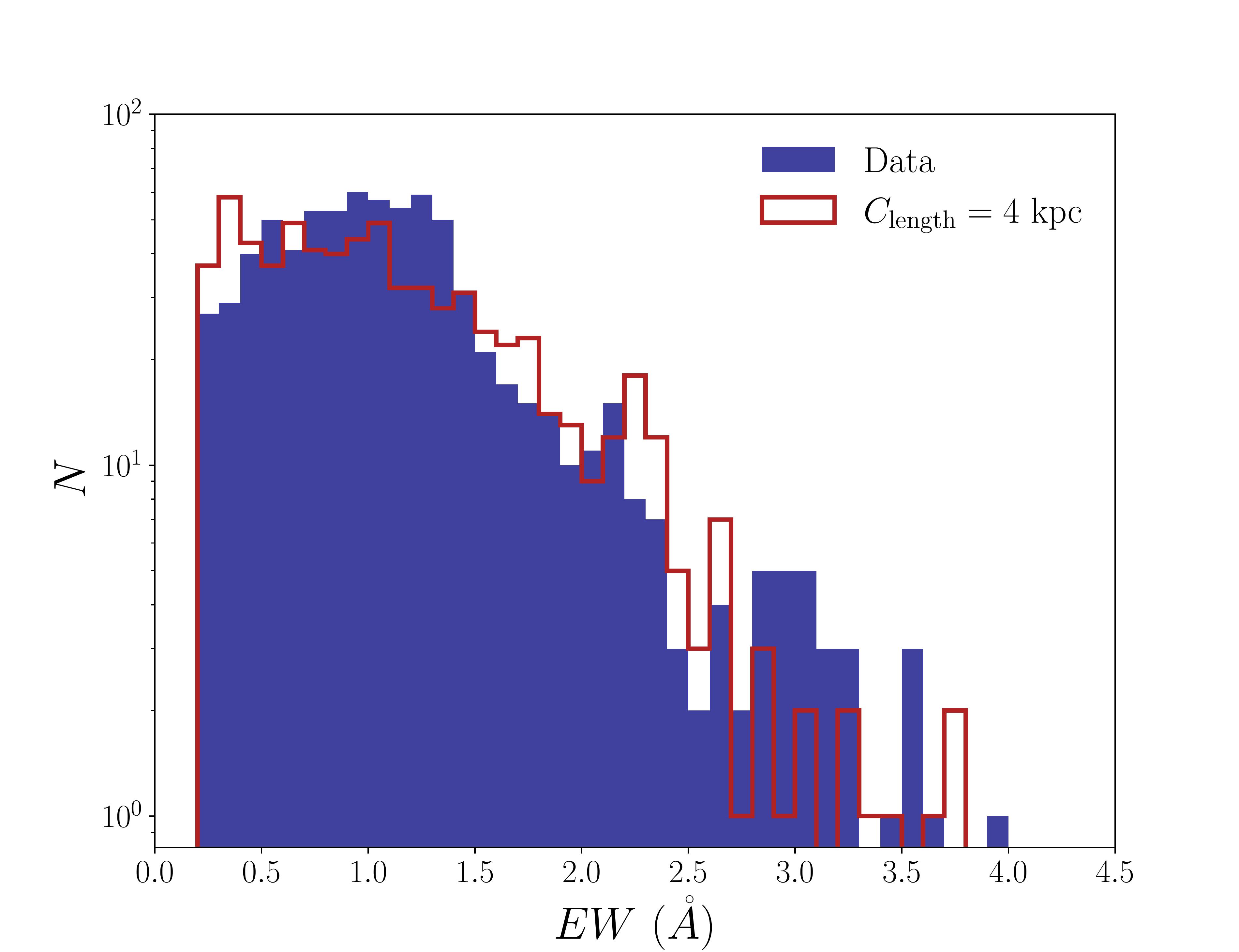}
   \caption{Distribution of \mgii\ EW in the data (blue) and models with $C_{\rm{length}}=4$ kpc (red). Both distributions are built by combining 10 different realizations of models and data (see Sect.~\ref{limitations}). Both data and models are incomplete for $\rm{EW}\lesssim1\ \AA$.}
              \label{fig:Completeness}%
    \end{figure}
Another main assumption of this work is the creation of our empirical models based on the quasar survey from \cite{huang21}, assuming that the cool CGM around our galaxies has properties similar to the galaxies of their sample. Our galaxies are at a median redshift $z\sim 0.8$, while those analyzed in \cite{huang21} are at a significantly lower redshift ($z\sim0.21$). During this cosmic time the properties of the cool circumgalactic medium might change and while we have seen that our data are consistent with the EW-vs-$R'$ relation found by \cite{huang21}, the intrinsic scatter of this relation could be different, affecting our results on the gas coherence length. 
However, there is to date no evidence of any dependence of this scatter across this range of redshifts. \cite{nielsen13} explored the relation between the \mgii\ absorption strength and the projected distance from the galaxy in the MAG{{\sc ii}}CAT sample, composed of almost 200 galaxies whose redshifts vary from 0.072 to 1.120. They found that there is no dependence with redshift in the scatter of this relation, which implicates that our results, although based on a sample of galaxies at lower redshift with respect to our data, are most likely robust. To further test this, we have run an additional Bayesian test on $C_{\rm{length}}$, leaving also the intrinsic scatter $\sigma$ used to create our models (see Sect.~\ref{mapCreation}) free to vary between 0.05 and 0.5. We find that the posterior distribution of $\sigma$ has a median at $0.282^{+0.175}_{-0.124}$ (2$\sigma$ uncertainties), interestingly very similar to the original value of 0.278 (based on the data from \citealt{huang21}) and that the findings on $C_{\rm{length}}$ (see Table~\ref{tab:testsCL}) do not change with respect to our fiducial model.

A limitation of this study is that our sample is composed of only three galaxies, whose CGM does not necessarily resemble the average best-fit relation of \cite{huang21}, which is based on hundreds of objects. The fact that our EWs  
lie on top of the \cite{huang21} relation seems to suggest that this is a good representation of the cool CGM of our galaxies, at least for $R'>10$ kpc; however, including a larger number of halos in our sample might increase the scatter in our data, potentially affecting our findings.
The tests described in Sect.~\ref{galvary} indicate that adding more galaxies to our sample (as long as they have similar properties to those analyzed here) might not influence our results, but a larger and more statistically significant sample of galaxies will be needed to confirm 
this. 
 
We have chosen to work with 
binned spaxels of $0.6\arcsec$ 
to maximize the number of probes, while still 
reducing the cross-talk between adjacent spaxels. The (slight) amount of cross-talk has the effect of producing smaller EW differences between adjacent spaxels and this might affect our findings. 
To evaluate the significance of this effect, we ran the same Bayesian analysis but using a $4\times4$ binning for the data (corresponding to $0.8\arcsec$, see also \citealt{lopez18,tejos21}), which further reduces the seeing-induced cross-correlation. The results,  reported in Table~\ref{tab:testsCL}, are consistent with our fiducial findings, with the median value of the posterior distribution well within the fiducial range of coherence lengths. The main difference is given by the fact that the posterior distribution is much wider (the 97.5 percentile is equal to almost 11 kpc), meaning that the constraints on the coherence scale are looser with respect to our fiducial case, due to the smaller number of spaxels. We therefore conclude that, as long as the binned spaxels have dimensions comparable to the observational seeing, our results are rather insensitive to the exact choice of spatial binning.

Finally, an inherent limitation of our dataset is given by its incompleteness 
at
$\rm{EW}\lesssim1\ \AA$. This can be seen in Fig.~\ref{fig:Completeness}, where we show in blue the EW distribution obtained by combining 10 different bootstrap realizations of the full data sample, drawing values from Gaussian distributions centered on the observed EWs and with a standard deviation equal to the observed uncertainties (as in the Bayesian analysis, see Sect.~\ref{Bayes}). We can see how, for EWs weaker than about $1\ \AA$, the data do not exhibit the exponential growth in the distribution 
expected for \mgii\ toward quasars 
\citep[see][]{zhu13}.  
On the other hand, 
the synthetic distribution 
appears also to be incomplete below $1$ \AA\
(red distribution of Fig.~\ref{fig:Completeness}). 
This is not surprising 
as we are excluding all the spaxels with non-detections in the data, which would yield a large number of low EW values in the model (being most of them at large impact parameters). 
The synthetic distribution was 
obtained by combining 10 different model realizations with a coherence length of 4 kpc, consistent with our estimated best-fit value. Even though the shape of this distribution slightly changes depending on the chosen $C_{\rm{length}}$, in all cases models and data are consistent with each other in the region of weak EWs. This feature holds also when looking at the three fields separately.

Would a higher S/N affect our results? Having spectra with a higher S/N would allow us to increase the number of detections in our data, especially at $R'\gtrsim50$ kpc, where most of our current data show only non-detections. While it is not trivial to predict how a more complete sample would affect our current findings, such a data-set would be crucial to more accurately study the possible variation of the coherence length with the distance from the central galaxy, since we would have a much better sampling at large impact parameters. 

To conclude, even though a higher S/N is desirable to conduct a more detailed analysis, S/N incompleteness does not affect the results of the current study. 

\section{Summary and conclusions}\label{conclusions}
In this paper, we investigated the coherence length ($C_{\rm{length}}$) of the cool circumgalactic medium traced by \mgii\ absorption. We utilized arc-tomographic MUSE data of the cool CGM of three different star-forming disk galaxies at redshift $\lesssim1$, which allowed us to have a sample of \mgii\ EWs coming from almost 100 spaxels spanning the galaxy halos up to several tens of kpc from the center. We developed 2D empirical models that describe the EW distribution and gas covering fraction, based on a recent QSO-survey of the cool CGM of low-redshift galaxies. We then compared models and data through a Bayesian analysis, in order to find the best coherence length that reproduces both the observed scatter in the distribution of EWs as a function of the impact parameter and the observed distribution of the fractional difference of EWs. Our main findings are the following:
\begin{enumerate}
\item our data are best reproduced by models with a cool gas coherence length in the range between 1.4 and 7.8 kpc (corresponding to the 2$\sigma$ limits of our posterior distribution), meaning that the \mgii\ absorption strength does not vary significantly below these scales;
\vspace{0.1cm}
\item we find that there is no significant trend of the coherence length with the impact parameter, possibly indicating similar formation mechanisms for the cool gas at different distances from the central galaxies. There is a slight tendency of having larger coherence lengths at larger impact parameters, but the variation is likely due to the low-number statistics and more data are needed to properly probe this trend;
\vspace{0.1cm}
\item we observe a small variation in the coherence lengths of the three individual galaxy halos, but this variation is well within our fiducial range. This suggests that similar galaxies are likely to be surrounded by CGM with a similar structure.
\end{enumerate}
To conclude, with this study we were able to put the most stringent constraints to date on the coherence scale of the cool CGM, thanks to the use of unique data that allowed us to spatially resolve this medium around individual galaxies at $z\sim1$. While here we made use of only three galaxy halos, a future larger sample of such data will possibly confirm and refine the findings of this paper. Moreover, while this analysis is entirely based on the strength of the \mgii\ absorption lines and on simple empirical models, future studies involving also the gas kinematics, the absorption from other ions and the use of more sophisticated and physically motivated models, will help in building a more complete picture of the CGM structure. Such results will be extremely useful to inform and constrain our current and future theoretical models of the gaseous halos of galaxies and to eventually understand how galaxies form and evolve in the Universe.

\begin{acknowledgements}
The authors would like to thank the referee for a constructive report. A.A. acknowledges the financial support of the Joint Committee ESO-Chile grant. S.L. acknowledges support by FONDECYT grant 1231187. E.J.J. acknowledges support from FONDECYT Iniciaci\'on en investigaci\'on 2020 Project 11200263 and the ANID BASAL project FB210003. L.F.B. acknowledges the support of FONDECYT project 23050313. 
\end{acknowledgements}
\bibliographystyle{aa} 
\bibliography{aanda}

\begin{appendix}
\section{Absorption line fits}\label{fits}
\begin{figure}[!h]
   \centering
   \includegraphics[clip, trim={0cm 0cm 0cm 0cm}, width=0.71\linewidth]{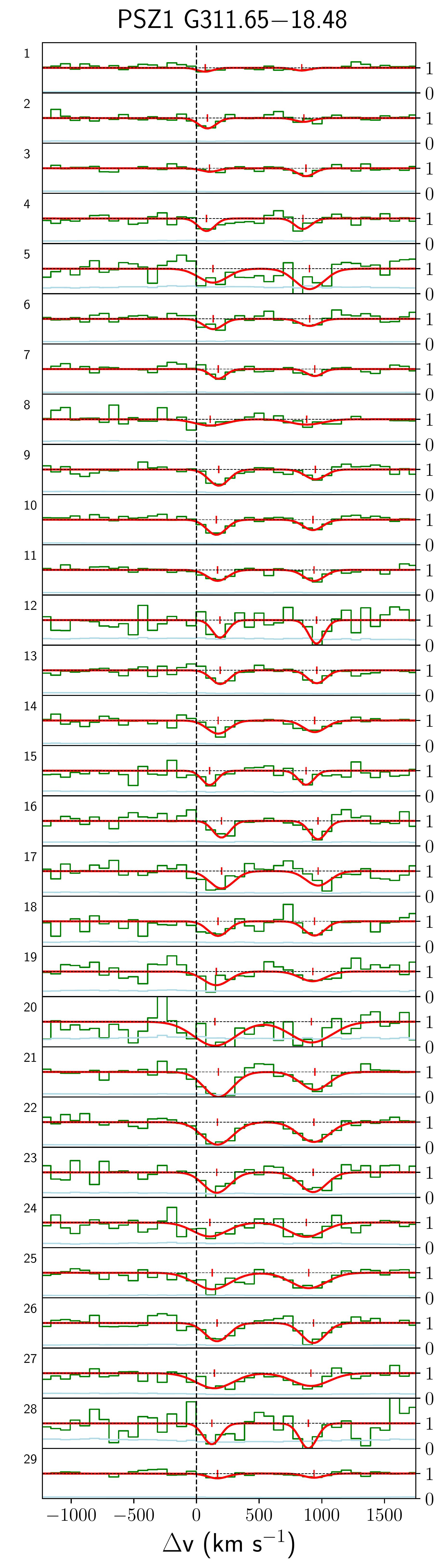}
   \caption{Results of the fitting analysis. Each panel shows a different spaxel (numbered arbitrarily) with \mgii\ detections in the \pks\ field. The normalized MUSE spectra are shown in green (uncertainties in light-blue). The Gaussian fits of the \mgii\ doublet are reported in red.}
              \label{fig:fitsPlanck}%
    \end{figure}

\begin{figure}
   \centering
   \includegraphics[clip, trim={0cm 0cm 0cm 0cm}, width=0.73\linewidth]{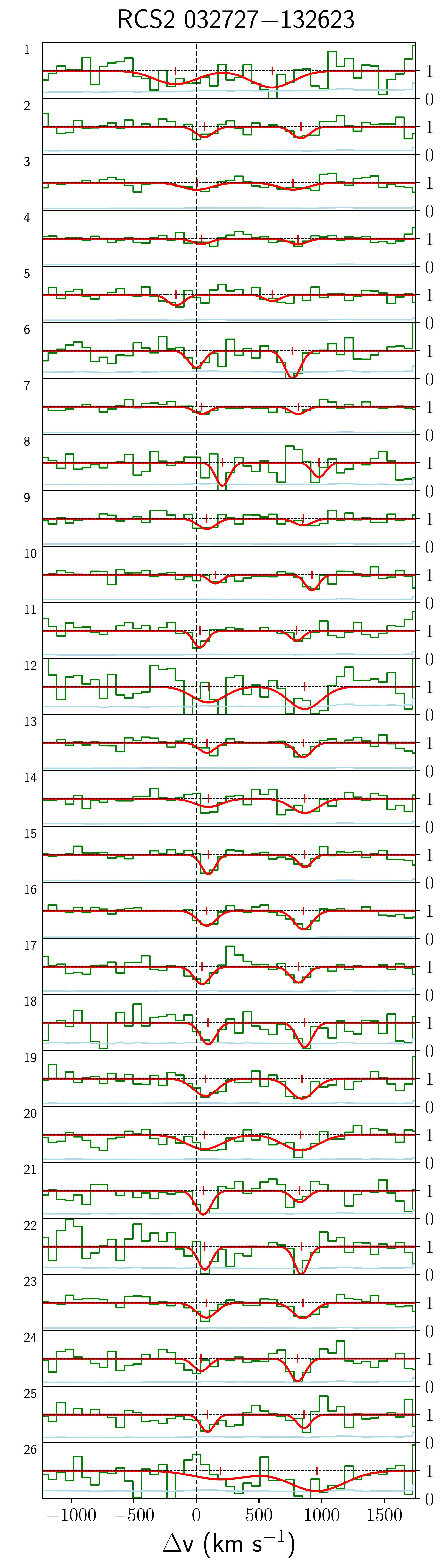}
   \caption{Same as Fig. \ref{fig:fitsPlanck}, for the \rcs\ field.}
              \label{fig:fitsRCS}%
    \end{figure}

\begin{figure}
   \centering
   \includegraphics[clip, trim={0cm 0cm 0cm 0cm}, width=0.73\linewidth]{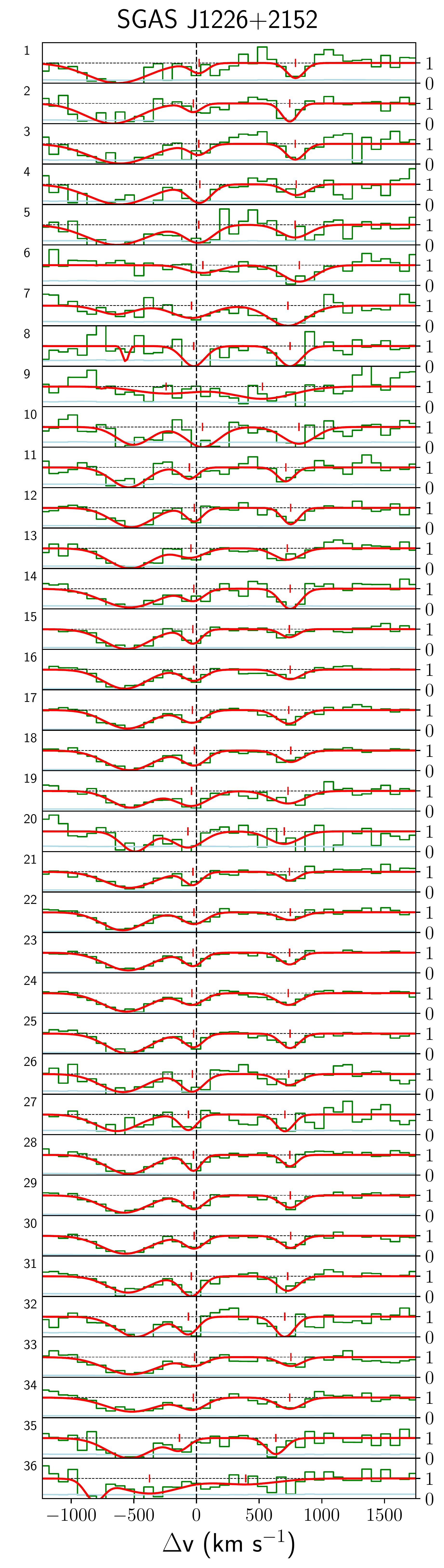}
   \caption{Same as Fig.~\ref{fig:fitsPlanck}, for the \sgas\ field. An additional Gaussian that accounts for the \siii\ $\lambda$1260 absorption is fitted.}
              \label{fig:fitsJ1226}%
    \end{figure}

In Tables~\ref{table_EWs_pks}, \ref{table_EWs_rcs} and \ref{table_EWs_sgas} and in Fig.s~\ref{fig:fitsPlanck}, \ref{fig:fitsRCS} and \ref{fig:fitsJ1226} we report the results of the automated Gaussian fits (Sect.~\ref{absorptionFits}) that were used to derive the EWs used in this work, for respectively the \pks, \rcs, and \sgas\ fields.
We show only the results concerning the successful fits, or detections (significance in the EW larger than 2 in both doublet lines), as the non-detections were excluded from our analysis. In the three figures, 
the normalized MUSE spectra are shown in green, with their respective uncertainties in blue, while the fits of the \mgii $\lambda\lambda 2796,2803$ doublet are reported in red, with the red ticks showing the positions of the line velocity centroids. Note that, for the \sgas\ field, the \mgii\ lines are partially blended with \siii$\lambda 1260$ absorption, so this is modeled by adding an additional Gaussian fit in our analysis for this field \citep[see][]{tejos21}.

For the reasons explained in Sect.~\ref{mainObs}, we do not consider all the spaxels with $R'\lesssim 10$ kpc\footnote{Note that in Tables~\ref{table_EWs_pks}--\ref{table_EWs_sgas} we are reporting the impact parameter $R$, not $R'$, which is re-scaled based on the galaxy mass (Eq.~\ref{eq:Rcorr}).}, therefore the spectra from number 21 to 28 in the \pks\ field and those from 6 to 10 and from 12 to 14 in the \sgas\ field are excluded from our analysis. The results of Sect.~\ref{results} are based on all the other spectra shown in Fig.s~\ref{fig:fitsPlanck}, \ref{fig:fitsRCS} and \ref{fig:fitsJ1226}. We note that, despite fulfilling the criteria of our automated fitting analysis, some of the Gaussian fits look `dubious' upon visual inspection. In particular, we consider poor fits those of the spectra 12 and 26 in the \rcs\ field and that of spectrum 36 in the \sgas\ field. To assess how much these `dubious' spectra influence our results, we ran an additional Bayesyan analysis (Sect.~\ref{Bayes}), excluding the EWs coming from the three spectra mentioned above. The resulting posterior distribution has a median $C_{\rm{length}}$ equal to $4.1\ \rm{kpc}$ and the 2$\sigma$ limits are equal to 1.3 and 7.7 kpc, therefore perfectly consistent with our fiducial findings. We conclude that this (small) set of uncertain fits does not affect the results of this paper.\\

\begin{table}[!h]
  \centering
\caption{Absorption line measurements for \pks.}
\begin{tabular}{cccccc}
\hline
\hline
Id.& $\Delta \alpha$ & $\Delta \delta$ & $R$ & EW & $\sigma_{\rm{EW}}$ \\
\vspace{0.2cm}
&(arcsec)&(arcsec)&(kpc)&(\AA) & (\AA)\\
1 & -7.29 & -8.94 & 43 & 0.22 & 0.09  \\ 
2 & -4.89 & -5.94 & 28 & 0.67 & 0.17  \\ 
3 & -4.29 & -5.34 & 25 & 0.23 & 0.1  \\ 
4 & -4.89 & -4.74 & 25 & 0.78 & 0.22  \\ 
5 & -2.49 & -4.14 & 18 & 1.5 & 0.6  \\ 
6 & -3.09 & -4.14 & 19 & 0.74 & 0.24  \\ 
7 & -3.69 & -4.14 & 20 & 0.56 & 0.13  \\ 
8 & -4.29 & -4.14 & 22 & 0.74 & 0.3  \\ 
9 & -2.49 & -3.54 & 16 & 1.23 & 0.21  \\ 
10 & -3.09 & -3.54 & 17 & 1.11 & 0.11  \\ 
11 & -3.69 & -3.54 & 19 & 0.84 & 0.17  \\ 
12 & -1.89 & -2.94 & 13 & 1.02 & 0.39  \\ 
13 & -2.49 & -2.94 & 14 & 0.94 & 0.15  \\ 
14 & -3.09 & -2.94 & 16 & 1.09 & 0.15  \\ 
15 & -3.69 & -2.94 & 17 & 0.83 & 0.28  \\ 
16 & -1.89 & -2.34 & 11 & 1.05 & 0.28  \\ 
17 & -2.49 & -2.34 & 13 & 1.56 & 0.31  \\ 
18 & -1.29 & -1.74 & 8 & 1.11 & 0.32  \\ 
19 & -1.89 & -1.74 & 9 & 1.32 & 0.56  \\ 
20 & 0.51 & -1.14 & 7 & 3.73 & 1.03  \\ 
21 & -0.09 & -1.14 & 6 & 2.68 & 0.29  \\ 
22 & -0.69 & -1.14 & 5 & 2.53 & 0.21  \\ 
23 & -1.29 & -1.14 & 6 & 1.99 & 0.34  \\ 
24 & 0.51 & -0.54 & 5 & 1.76 & 0.45  \\ 
25 & -0.09 & -0.54 & 3 & 2.19 & 0.26  \\ 
26 & -0.69 & -0.54 & 3 & 1.61 & 0.24  \\ 
27 & -0.09 & 0.06 & 0 & 2.04 & 0.45  \\ 
28 & -0.69 & 0.06 & 3 & 1.34 & 0.48  \\ 
29 & 6.51 & 4.26 & 29 & 0.35 & 0.11  \\ 
\label{table_EWs_pks}
\end{tabular} 
\flushleft
Notes: $\Delta \alpha$ and $\Delta \delta$ indicate the relative spaxel position with respect to G1 in the image plane, $R$ is the impact parameter in the absorber plane, and EW and $\sigma_{\rm EW}$ correspond to the \mgii$\lambda 2796$ rest-frame equivalent width and its 1-$\sigma$ uncertainty, respectively.
\end{table}

\begin{table}[!h]
  \centering
\caption{Same as Table~\ref{table_EWs_pks}, for \rcs.}
\begin{tabular}{cccccc}
\hline
\hline
Id.& $\Delta \alpha$ & $\Delta \delta$ & $R$ & EW & $\sigma_{\rm{EW}}$ \\
\vspace{0.2cm}
&(arcsec)&(arcsec)&(kpc)&(\AA) & (\AA)\\
1 & 2.77 & -11.09 & 52 & 1.89 & 0.69  \\ 
2 & 3.37 & -10.49 & 50 & 0.61 & 0.26  \\ 
3 & 3.97 & -9.89 & 47 & 0.69 & 0.23  \\ 
4 & 3.97 & -9.29 & 45 & 0.32 & 0.15  \\ 
5 & 3.37 & -9.29 & 45 & 0.59 & 0.19  \\ 
6 & 3.37 & -6.89 & 35 & 0.88 & 0.38  \\ 
7 & 2.77 & -6.89 & 36 & 0.37 & 0.14  \\ 
8 & 0.37 & -5.69 & 33 & 1.05 & 0.4  \\ 
9 & 0.97 & -5.09 & 29 & 0.62 & 0.22  \\ 
10 & 0.37 & -5.09 & 30 & 0.4 & 0.17  \\ 
11 & 0.37 & -4.49 & 27 & 0.76 & 0.27  \\ 
12 & -0.83 & -4.49 & 29 & 1.72 & 0.69  \\ 
13 & -1.43 & -3.29 & 24 & 0.54 & 0.15  \\ 
14 & -2.03 & -3.29 & 25 & 0.73 & 0.21  \\ 
15 & -1.43 & -2.69 & 20 & 0.9 & 0.15  \\ 
16 & -2.03 & -2.69 & 22 & 0.94 & 0.09  \\ 
17 & -2.63 & -2.69 & 24 & 0.95 & 0.23  \\ 
18 & -1.43 & -2.09 & 17 & 1.11 & 0.46  \\ 
19 & -2.03 & -2.09 & 19 & 1.4 & 0.26  \\ 
20 & -2.63 & -2.09 & 21 & 1.74 & 0.31  \\ 
21 & -3.23 & -2.09 & 23 & 1.25 & 0.37  \\ 
22 & -4.43 & -2.09 & 27 & 1.04 & 0.36  \\ 
23 & -4.43 & -1.49 & 25 & 1.01 & 0.18  \\ 
24 & -5.03 & -1.49 & 27 & 0.63 & 0.24  \\ 
25 & -5.63 & -0.89 & 27 & 0.78 & 0.36  \\ 
26 & -9.83 & -0.89 & 44 & 1.57 & 0.72  \\ 
\label{table_EWs_rcs}
\end{tabular} 
\end{table}

\begin{table}[!h]
  \centering
\caption{Same as Table~\ref{table_EWs_pks}, for \sgas.}
\begin{tabular}{cccccc}
\hline
\hline
Id.& $\Delta \alpha$ & $\Delta \delta$ & $R$ & EW & $\sigma_{\rm{EW}}$ \\
\vspace{0.2cm}
&(arcsec)&(arcsec)&(kpc)&(\AA) & (\AA)\\
1 & 2.15 & -3.08 & 17 & 0.78 & 0.31  \\ 
2 & 1.55 & -3.08 & 17 & 0.64 & 0.29  \\ 
3 & 0.95 & -3.08 & 17 & 0.92 & 0.41  \\ 
4 & -0.85 & -2.48 & 15 & 1.89 & 0.48  \\ 
5 & -1.45 & -2.48 & 16 & 2.37 & 0.77  \\ 
6 & 0.35 & -0.08 & 1 & 1.18 & 0.46  \\ 
7 & -0.85 & -0.08 & 4 & 2.19 & 0.64  \\ 
8 & 0.35 & 0.52 & 3 & 2.11 & 0.64  \\ 
9 & -0.25 & 0.52 & 3 & 1.96 & 0.76  \\ 
10 & -0.85 & 0.52 & 4 & 3.0 & 0.79  \\ 
11 & 0.95 & 1.72 & 10 & 0.91 & 0.3  \\ 
12 & 0.35 & 1.72 & 9 & 1.05 & 0.19  \\ 
13 & -0.25 & 1.72 & 8 & 1.13 & 0.25  \\ 
14 & -0.85 & 1.72 & 8 & 1.06 & 0.29  \\ 
15 & 0.95 & 2.32 & 13 & 1.15 & 0.2  \\ 
16 & 0.35 & 2.32 & 12 & 1.14 & 0.12  \\ 
17 & -0.25 & 2.32 & 11 & 1.37 & 0.1  \\ 
18 & -0.85 & 2.32 & 11 & 1.62 & 0.14  \\ 
19 & -1.45 & 2.32 & 12 & 2.28 & 0.35  \\ 
20 & -2.05 & 2.32 & 13 & 2.3 & 0.77  \\ 
21 & 0.95 & 2.92 & 15 & 1.04 & 0.21  \\ 
22 & 0.35 & 2.92 & 14 & 1.3 & 0.15  \\ 
23 & -0.25 & 2.92 & 14 & 1.16 & 0.08  \\ 
24 & -0.85 & 2.92 & 14 & 1.27 & 0.1  \\ 
25 & -1.45 & 2.92 & 14 & 1.2 & 0.12  \\ 
26 & -2.05 & 2.92 & 15 & 1.87 & 0.4  \\ 
27 & 0.95 & 3.52 & 18 & 1.25 & 0.4  \\ 
28 & -0.25 & 3.52 & 16 & 1.11 & 0.15  \\ 
29 & -0.85 & 3.52 & 16 & 1.32 & 0.13  \\ 
30 & -1.45 & 3.52 & 17 & 1.19 & 0.14  \\ 
31 & -2.05 & 3.52 & 17 & 1.92 & 0.31  \\ 
32 & -0.25 & 4.12 & 19 & 1.56 & 0.42  \\ 
33 & -0.85 & 4.12 & 19 & 0.85 & 0.23  \\ 
34 & -1.45 & 4.12 & 19 & 1.37 & 0.27  \\ 
35 & -2.05 & 4.12 & 19 & 1.01 & 0.38  \\ 
36 & -1.45 & 4.72 & 21 & 4.2 & 1.27  \\ 
\label{table_EWs_sgas}
\end{tabular} 
\end{table}

\end{appendix}
%-------------------------------------------------------------------

\end{document}